\documentclass[11pt]{article}

\pdfoutput=1

\usepackage[makeroom]{cancel}
\usepackage{color}
\usepackage{graphicx}
\usepackage{amsmath}
\usepackage{amssymb}
\usepackage{xspace}
\usepackage[small]{subfigure}
\usepackage[numbers,compress]{natbib}
\usepackage[hyperfootnotes=false]{hyperref}
\usepackage{tcolorbox}

\newlength{\fighskip} \fighskip=2pt
\newlength{\figvskip} \figvskip=3pt

\newcommand*{\figbox}[2]{{
  \def\figscale{#1}
  \def\arraystretch{0.8}
  \arraycolsep=0pt
  \begin{array}{c}
    \vbox{\vskip\figscale\figvskip
      \hbox{\hskip\figscale\fighskip
        \includegraphics[scale=\figscale]{#2}}}
  \end{array}}}

\DeclareMathOperator{\Tr}{Tr}
\DeclareMathOperator{\Pauli}{Pauli}
\DeclareMathOperator{\Prob}{Prob}
\DeclareMathOperator{\Sum}{Sum}
\DeclareMathOperator{\ext}{ext}
\DeclareMathOperator{\Stab}{Stab}
\DeclareMathOperator{\Logic}{Logic}
\DeclareMathOperator{\rev}{rev}
\DeclareMathOperator{\Bell}{Bell}

\DeclareMathOperator{\Comm}{Comm}
\DeclareMathOperator{\C}{\mathcal{C}}
\DeclareMathOperator{\E}{\mathcal{E}}

\usepackage{mciteplus} 
\usepackage{dcolumn}
\usepackage{bm}
\usepackage{verbatim}
\usepackage{amscd}
\usepackage{amsfonts}
\usepackage{setspace}
\usepackage{amsthm}
\usepackage{enumerate}
\usepackage{mathtools}

\theoremstyle{plain}
\newtheorem{theorem}{Theorem}
\theoremstyle{plain}
\newtheorem{lemma}{Lemma}
\theoremstyle{plain}
 
\theoremstyle{plain}

\theoremstyle{remark}

\theoremstyle{conjecture}

\theoremstyle{corollary}
\newtheorem{corollary}{Corollary}

\begin{document}

\title{\bf 
Decoding the Entanglement Structure of \\ Monitored Quantum Circuits
}
\author{
Beni Yoshida\\ 
{\em \small Perimeter Institute for Theoretical Physics, Waterloo, Ontario N2L 2Y5, Canada} }
\date{}

\maketitle

\begin{abstract}

Given an output wavefunction of a monitored quantum circuit consisting of both unitary gates and projective measurements, we ask whether two complementary subsystems are entangled or not. For Clifford circuits, we find that this question can be mapped to a certain classical error-correction problem where various entanglement measures can be explicitly computed from the recoverability. The dual classical code is constructed from spacetime patterns of out-of-time ordered correlation functions among local operators and measured Pauli operators in the past, suggesting that the volume-law entanglement in a monitored circuit emerges from quantum information scrambling, namely the growth of local operators. 
We also present a method of verifying quantum entanglement by providing a simple deterministic entanglement distillation algorithm, which can be interpreted as decoding of the dual classical code. Discussions on coding properties of a monitored Clifford circuit, including explicit constructions of logical and stabilizer operators, are also presented. 
Applications of our framework to various physical questions, including non-Clifford systems, are discussed as well. Namely, we argue that the entanglement structure of a monitored quantum circuit in the volume-law phase is largely independent of the initial states and past measurement outcomes except recent ones, due to the decoupling phenomena from scrambling dynamics, up to a certain polynomial length scale which can be identified as the code distance of the circuit. We also derive a general relation between the code distance and the sub-leading contribution to the volume-law entanglement entropy. Applications of these results to black hole physics are discussed as well. 

\end{abstract}

\newpage

\tableofcontents

\vspace{-0.7\baselineskip}

\newpage

\section{Introduction}

Recently it has been discovered that monitored quantum circuits consisting of both interacting unitary dynamics and local projective measurements can retain long-range entanglement obeying the volume-law~\cite{Li:2018aa, Skinner:2019aa}. These theoretical findings hint far-reaching possibility that quantum entanglement may play crucial roles in the physics of many-body quantum systems outside controlled laboratory setups where the systems are continuously monitored by observers and decohere to the environment. Indeed, it is illuminating to remind ourselves that objects surrounding our daily lives, such as a cup of coffee, are after all quantum many-body systems which evolve unitarily in the presence of continuous measurements. However, if entanglement in monitored quantum systems would ever be relevant to naturally occurring and observable physical phenomena, the entanglement must be verifiable by some simple physical processes since observing such phenomena would verify the entanglement. While previous studies on monitored quantum circuits have revealed interesting features of entanglement phase transitions driven by measurement rates (see~\cite{Ippoliti:2021ab, Jian:2020aa, Lavasani:2021aa, Zabalo:2020aa, Szyniszewski:2019aa, Sang:2021aa, Li2020, Nahum:2021aa, Vijay2020, Zhang:2020aa, Tang:2020aa, Bao2020} for samples of previous works), our current understanding of the \emph{entanglement structure} arising in a monitored quantum circuit remains elusive with no known universal method of verifying quantum entanglement. 

In this paper, we investigate the entanglement structure arising in a monitored quantum circuit. We will pay particular attention to the following three key questions.
\begin{enumerate}[(a)]
\item \textbf{Entanglement Structure:} Given an output wavefunction of a monitored quantum circuit, how is a subsystem $A$ entangled with its complementary subsystem $B$? 
\item \textbf{Entanglement distillation:} When two subsystems $A$ and $B$ are entangled with each other, how do we verify their entanglement? Specifically, how do we distill simple entangled states (such as EPR pairs) from $A$ and $B$?
\item \textbf{Measurement dependence
~\footnote{The entanglement structure of a monitored quantum circuit depends on the measurement outcomes in the past, as well as the initial states of the circuit, until these are forgotten after an exponentially long time-evolution. One might then expect that verifying the entanglement requires knowledge of measurement outcomes in the distant past. The nature, however, would not be keeping a record of exponentially many measurement outcomes and utilize them cleverly to verify the entanglement. Hence, if the entanglement arising in a monitored many-body quantum system is to be relevant to some observable phenomena, it should \emph{not} depend on measurement outcomes in the distant past or the initial states. In this paper, we will argue that this is indeed the case below a certain length scale.}
:} How does the entanglement structure of a monitored quantum circuit depend on measurement results in the past? To what extent do measurement outcomes in the past influence the entanglement structure? Relatedly, does the entanglement depend on the initial states of the circuit?
\end{enumerate}

In this paper, we will address these questions by focusing on monitored quantum circuits whose unitary part of the dynamics are supplied by Clifford operators, which are unitary operators that transform Pauli operators to (possibly different) Pauli operators. While Clifford dynamics differs from generic dynamics of interacting many-body quantum systems in crucial ways, Clifford dynamics can teach us qualitative features of entanglement structure that are universal for monitored quantum circuits. Our goal is to develop a theoretical tool to understand the entanglement structure arising in a monitored Clifford circuit and propose a simple entanglement distillation algorithm that verifies quantum entanglement between two subsystems $A$ and $B$. Building on these results on monitored Clifford circuits, we will obtain some physical implications which can be applied widely to generic monitored quantum circuits. 

\subsection{Previous works}

The central challenge is to reveal the entanglement structure, namely to understand how two subsystems are entangled in the output wavefunction of a monitored circuit. 
This question can be addressed unambiguously by solving the \emph{entanglement distillation problem}. Loosely speaking, if two subsystems $A$ and $B$ are entangled with each other, one should be able to distill quantum entanglement between $A$ and $B$ and convert it into some ``usable'' or ``simple'' forms of entangled states, such as an EPR pair $\frac{1}{\sqrt{2}} (|00\rangle + |11\rangle$), by acting only on $A$ and $B$ locally. Entanglement distillation typically requires us to localize the entangled degrees of freedom on $A$ and $B$ into locally supported qubits. This is what we mean by understanding and verifying the entanglement structure~\footnote{One might think of preparing two copies of the output wavefunctions and measure the R\'{e}nyi-$2$ entropy. But finding the (naive R\'{e}nyi-$2$ generalization of) mutual information, for instance, requires us to find $S_{B}^{(2)}$ by measuring $\Tr(\rho_{B}^2)$ which will be exponentially small in most of the interesting cases. In addition, preparing identical copies will be even more difficult for monitored systems since measurement outcomes in two copies must be identical as well.}.

One possible approach toward the entanglement verification is to interpret the entanglement distillation as a decoding problem and use the Pets recovery map by viewing the output wavefunction as a quantum channel from $A$ to $B$ via the Choi-Jamio\l{}kowski isomorphism~\cite{Ohya_Petz_Text, in_prep}. However, the Petz map is a quantum operation that does not necessarily have simple physical realizations. Indeed, its physical implementation typically requires post-selection or amplitude amplification (\emph{i.e.} use of the Grover search algorithm) which may not be physically simple or computationally efficient, especially when the subsystem $A$ becomes large~\cite{Yoshida:2017aa}. 

Another interesting approach toward characterization of the volume-law entanglement is to interpret a monitored quantum circuit as a quantum error-correcting code~\cite{Choi:2020aa, Gullans:2020aa, Gullans:2020ab, Fan:2021aa, Li:2021aa}. Namely, instead of starting from a pure state, maximally mixed states are prepared as the initial states. By purifying the system with an ancilla reference system $R$ which is entangled with the original system, the circuit can be viewed as a quantum error-correcting code where the system stores quantum information as entanglement between the system and the reference. The key observation is that the volume-law entanglement is protected from local projective measurements via quantum error-correction~\cite{Choi:2020aa}. Recent studies have also numerically verified that entanglement phase transition can be addressed by studying the coding properties of a monitored quantum circuit. 

Despite its conceptual novelty, the quantum error-correction approach has a crucial drawback of not being a direct measure of quantum entanglement arising in a monitored quantum system itself. Indeed, it remains puzzling why the entanglement between the system and the reference $R$ may serve as a probe of the entanglement within the system. Verification of the entanglement between the system and the reference is also a non-trivial task. Another issue is that the quantum memory, stored in a monitored circuit, will be eventually lost after an exponentially long time-evolution~\cite{Gullans:2020aa, Fidkowski:aa}. Yet, the volume-law entanglement from a monitored quantum circuit remains even after the circuit loses its initial quantum information. Here we hope to understand universal signatures of the entanglement structure in a monitored quantum circuit which is independent of the reference system $R$ and is applicable at any given moment, including moments after an exponentially long time-evolution.

Another interesting approach is to simulate a monitored circuit by a unitary circuit without measurements via a certain spacetime duality~\cite{Ippoliti:2021aa}. 

\begin{figure}
\centering
(a)\includegraphics[width=0.45\textwidth]{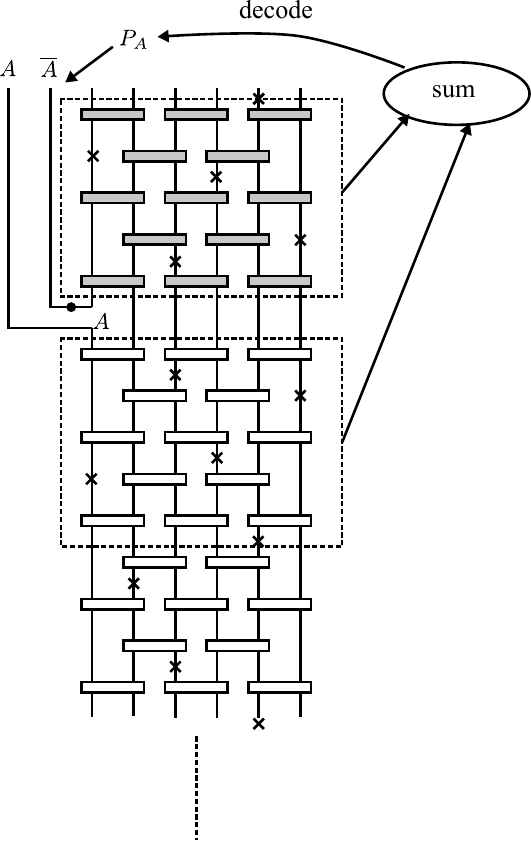} \qquad
(b)\includegraphics[width=0.32\textwidth]{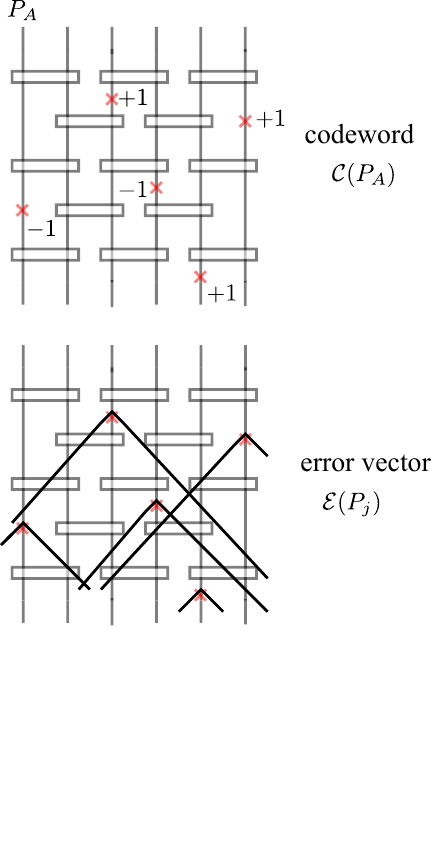}
\caption{(a) A summary of the entanglement distillation algorithm. Given an output wavefunction of a monitored Clifford circuit, we insert additional EPR pairs on $A$ and $\overline{A}$ (shown as a horizontal line with a black dot). We then implement the same measurement sequence in a reverse order (shown in shaded blocks). A sum of the original and reverse measurement results $m$ and $\overline{m}$ generates a bit string $s = m \cdot \overline{m}$. Based on the bit string $s$, we apply some feedback Pauli operator $P_{A}$. An appropriate feedback operation can be found by error-correcting this bit string $s$ into a codeword bit string $\C(P_{A})$.  As shown in the figure, one needs to reverse only a part of the original monitored circuit since measurement histories in the distant past will not influence the entanglement between $A$ and $B$ due to the decoupling phenomena arising from scrambling dynamics. 
(b) Construction of a dual classical error-correcting code. Here the codeword $\C(P_{A})$ records the space-time pattern of the operator growth of $P_{A}$ as OTOCs with respect to local Pauli operators which were projectively measured in the past. Error vectors $\E(P_{j})$ corresponds to OTOCs between a measured Pauli operator $P_{j}$ and other measured Pauli operators in the past. 
}
\label{fig-idea}
\end{figure}

\subsection{Main results}

\subsubsection{Entanglement structure (Section~\ref{sec:classical_code},~\ref{sec:distillation},~\ref{sec:scrambling})}

In this paper, we develop a theoretical framework to investigate the entanglement structure arising in a monitored Clifford circuit and present an entanglement distillation algorithm that verifies quantum entanglement in two complementary subsystems. The main results are summarized as follows. (See Fig.~\ref{fig-idea}).

\begin{enumerate}[(a)]
\item \textbf{Dual classical code problem:} We will show that the problem of revealing the entanglement structure of a monitored Clifford circuit can be mapped to a certain \emph{classical error-correction problem} where the recoverability of initial classical information corresponds to the presence of entanglement between two subsystems $A$ and $B$. 
\item \textbf{Entanglement distillation algorithm:} We will present a simple deterministic algorithm to distill EPR pairs from two complementary subsystems $A$ and $B$. The algorithm can be interpreted as a decoding procedure of the dual classical error-correcting code.
\end{enumerate}

We will begin by showing that a certain dual classical error-correction problem can be employed to study the entanglement structure of a monitored Clifford circuit. The corresponding classical code is constructed by examining commutation relations among local Pauli operators on a subsystem $A$ and measured Pauli operators $P_{j}$ in the past. Given a Pauli operator $P_{A}$ on a subsystem $A$, we think of encoding $P_{A}$ into a codeword vector $\C(P_{A})$ by recording its commutation relations with respect to measured Pauli operators $P_{j}$. These codeword vector $\C(P_{A})$ will be acted by error vectors $\E(P_{j})$ which account for commutation relations among measured Pauli operators $P_{j}$'s in a certain manner so that causal orderings are taken into account. See Fig.~\ref{fig-idea}(b). The central result is that two subsystems $A$ and $B$ are maximally entangled if and only if the initial information $P_{A}$ can be recovered even when error vectors $\E(P_{j})$ act on codeword vectors $\C(P_{A})$. In other words, the recoverability of the dual classical error-correcting code serves as a necessary and sufficient condition for maximal quantum entanglement between $A$ and $B$. In fact, by studying how much of classical information remains recoverable, one can explicitly compute the conditional entropy $S_{A|B}$:
\begin{align}
S_{A|B} \equiv S_{AB} - S_{B}
\end{align}
where the recoverability of the classical code corresponds to the negativity of the conditional entropy. The conditional entropy can be interpreted as the coherent quantum information when we view the output wavefunction as a quantum channel from $A$ to $B$. As such, the recoverability of the dual classical code underpins the robustness of quantum entanglement in a monitored Clifford circuit. 

We then present a deterministic algorithm for distilling quantum entanglement from $A$ and $B$. The algorithm implements the reverse of the monitored Clifford circuit as shown in Fig.~\ref{fig-idea}. When the measurement outcomes are ``favorable'', EPR pairs will be automatically distilled without the need of further actions. When the measurement outcomes are not ``favorable'', then some feedback operation is needed. The appropriate feedback operation can be found by solving the decoding problem of the dual classical code. Specifically, letting $m$ and $\overline{m}$ be the vectors which record measurement outcomes in the original circuit and the reverse circuit respectively, the sum vector $s = m\cdot \overline{m}$ plays the central role in the distillation algorithm. Namely, the sum vector $s$ is interpreted as an outcome of applying some error vectors on codeword vectors. By decoding the sum vector $s$, one can recover the original classical information which corresponds to some Pauli operator $P_{A}$. This $P_{A}$ is the necessary feedback operator to distill EPR pairs. 

We also present an application of this distillation algorithm to a certain proposal by Gullans and Huse which aims at detecting the entanglement phase transition by entangling the system of a monitored Clifford circuit to a single qubit (or a few qubits)~\cite{Gullans:2020ab}. 

It is well known that, due to the Gottesman-Knill theorem, Clifford circuits can be decoded efficiently. Here it is worth emphasizing that our algorithm is more efficient than these generic treatments.

These findings suggest that entanglement in a monitored quantum circuit emerges from \emph{scrambling dynamics}, namely the growth of local operators on $A$ by backward time-evolution which overlaps non-trivially with measured local operators in the past. Indeed, encoding into codewords of a dual classical error-correcting code can be interpreted as a space-time pattern of the operator growth (or the out-of-time ordered correlation (OTOC) functions). Hence, our results provide a rigorous and concrete argument to support the folklore belief that the scrambling dynamics, in a sense of OTOC functions, is necessary for the emergence of the volume-law entanglement phase in monitored quantum circuits.

\subsubsection{Coding properties (Section~\ref{sec:reference},~\ref{sec:R-distillation}, and Appendix~\ref{app:code})}

We will also study the coding properties of a monitored Clifford circuit by interpreting it as a quantum error-correcting code entangled with the reference system $R$. The framework of using a dual classical error-correcting code enables us to study the entanglement between the system and the reference $R$ as well. The main results are summarized as follows. 

\begin{enumerate}[(a)]
\item \textbf{Entanglement between system and reference:} We will derive explicit formulae of entanglement entropies for subsystems involving $A$, $B$, and $R$. We will also present an algorithm to distill an entangled state from the system and the reference $R$, and show that it is identical to the Choi-Jamio\l{}kowski state of a monitored Clifford circuit viewed as a stabilizer code. 
\item \textbf{Stabilizer and logical operators:} We will present explicit constructions of stabilizer and logical operators by using the dual classical error-correcting code. We will also derive a version of the cleaning lemma for monitored Clifford circuits. 
\end{enumerate}

To study the entanglement structure involving the reference, we will utilize the formula for the conditional entropy by viewing $A$, $B$, and the whole system $AB$ as input subsystems of the dual code. In this analysis, Pauli operators which become \emph{indistinguishable} from the identity operator play a crucial role:
\begin{align}
\mathcal{L} \equiv \big\langle P \in \Pauli : \C(P) \in \E  \big\rangle,\qquad \E \equiv \big\langle \{ \E(P_{j}) \}_{\forall j} \big\rangle.
\end{align}
Such Pauli operators will be referred to as \emph{null operators}. We find that entanglement entropies in subsystems can be written simply in terms of the numbers of null operators. For instance, the mutual information is given by 
\begin{align}
I_{(A,B)} = \log \frac{N_{I}}{N_{I_{A}}N_{I_{B}}}
\end{align}
where $N_{I_{A}}, N_{I_{B}}, N_{I}$ represent the numbers of null operators supported on $A$, $B$, and the whole system $AB$ respectively.

It turns out that the null operators of the dual classical code play the role of \emph{logical operators}. We will prove this statement by presenting an explicit recipe of recursively constructing stabilizer operators from measured Pauli operators $P_{j}$. 

We will also present an algorithm to distill an entangled state between the system $AB$ and the reference $R$. While the algorithm is simple, finding an appropriate feedback operator requires extra caution. In the dual classical code, the error vectors $\E(P_{j})$ were constructed by examining commutation relations with other Pauli operators $P_{i}$ in the past ($i<j$). Here, in order for the entanglement distillation between the system and the reference, we will need to construct the error vectors $\E_{\rev}(P_{j})$ in a \emph{reverse chronological order}, namely by examining commutation relations with respect to other Pauli operators $P_{i}$ in the \emph{future} ($i>j$). The algorithm generates the Choi-Jamio\l{}kowski state of the corresponding stabilizer code, confirming the quantum error-correcting code interpretation of a monitored Clifford circuit. 

\subsubsection{Hierarchy of entanglement structure (Section~\ref{sec:state-independence},~\ref{sec:state-dependence})}

\begin{figure}
\centering
\includegraphics[width=0.4\textwidth]{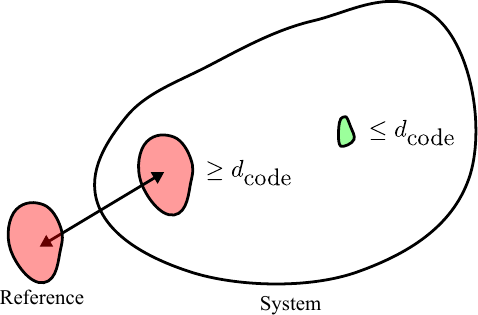} 
\caption{A cartoon of the hierarchy of the entanglement structure. A subsystem smaller than the code distance $d_{\text{code}}$ is entangled within the system, and this entanglement does not depend on the measurement outcomes in the distant past or the initial states of the circuit. A subsystem larger than $d_{\text{code}}$ is entangled with the reference system as well, and this entanglement is state-dependent.
}
\label{fig-Hierarchy}
\end{figure}

Our results reveal a certain interesting feature of the entanglement structure of a monitored quantum circuit in the volume-law phase. We will argue that the entanglement structure changes drastically when the subsystem $A$ exceeds a certain polynomial size scale that can be identified as the code distance of the circuit (Fig.~\ref{fig-Hierarchy}).

\begin{enumerate}[(a)]
\item \textbf{Below the code distance scale:} The entanglement between $A$ and its complement $B$ is independent of the initial states of the circuit. Furthermore, the entanglement does not depend on measurements that occurred more than the entanglement equilibrium time before. 
\item \textbf{Above the code distance scale:} The entanglement between $A$ and $B$ depends on the initial states as well as measurement outcomes in the distant past. Nevertheless, the value of the entanglement entropy $S_{A}$ does not depend on the choice of the initial states once the system reaches the entanglement equilibrium. 
\end{enumerate}

Our argument is based on a simple observation based on the decoupling phenomena. We expect that the monitored quantum circuit in the volume-law phase will reach the entanglement equilibrium in the $O(L)$ time scale ($L$ being the linear length), and the entanglement with the reference remains stable until an exponentially long quantum memory time. In the entanglement equilibrium, a subsystem $A$ smaller than the code distance will be decoupled from the reference system $R$, satisfying $I(A,R)\approx 0$. This suggests that any quantum operation acting on $R$ cannot influence the entanglement between $A$ and $B$. Observe that projecting the reference $R$ onto a product state $|0\rangle^{\otimes n}$ will set the initial state of the circuit as $|0\rangle^{\otimes n}$. Even after this projection, two subsystems $A$ and $B$ should remain entangled in the same manner. Hence, the entanglement structure below the code distance scale is \emph{state-independent}. Furthermore, since the decoupling of $A$ and $R$ occurs in the entanglement equilibrium time, the entanglement between $A$ and $B$ depend only on recent measurement outcomes up to the entanglement equilibrium time in the past. 

Above the code distance scale, we will have $I(A,R)\gtrapprox 0$ and hence, the entanglement between $A$ and $B$ will be \emph{state-dependent}. The entanglement verification requires knowledge of measurement outcomes in the distant past as well as the initial state, and is expected to be computationally intractable. Nevertheless, we expect that the value of the entanglement entropy $S_{A}$ will remain independent of the choice of the initial states. Indeed, one can explicitly show that $S_{A}$ does not change (except small statistical fluctuations) by choosing a Haar random initial state. Namely, the random projection on $R$ lets the entanglement between $A$ and $R$ join the entanglement between $A$ and $B$. In the language of quantum information theory, this mechanism is akin to the \emph{entanglement swapping} (or the quantum teleportation) driven by a random projection. As such, the volume-law behavior $S_{A}\approx a |A|$ persists across the code distance scale regardless of the choice of the initial states even though the nature of the entanglement structure changes drastically.

In order for a monitored quantum circuit to have an exponential quantum memory time, the code distance should scale polynomially with respect to the system size $n$. As such, the entanglement structure undergoes a transition from being state-independent to being state-dependent at an ``intermediate'' length scale. We will argue that the above observations can be supported on generic grounds for non-Clifford circuits as well. 

\subsubsection{Other applications (Section~\ref{sec:gamma},~\ref{sec:BH})}

Based on the aforementioned results, we will address two concrete physical questions concerning monitored quantum circuits.

\begin{enumerate}[(a)]

\item \textbf{Sub-leading contribution:} 
We will derive a general relation between coding properties of one-dimensional monitored quantum circuits and the sub-leading contribution to the volume-law entanglement entropy. Namely, we will show that, if the code distance scales as $d_{\text{code}}\approx n^{\gamma_{\text{code}}}$, then the entanglement entropy must scale as $S_{A} \approx a n_{A} + b n_{A}^{\gamma}$ with $\gamma=\gamma_{\text{code}}$.
 
\item \textbf{Relation to black hole physics:} We will argue that a monitored quantum circuit can be interpreted as the Hayden-Preskill recovery problem, running backward in time, where the late Hawking radiations are sequentially measured projectively. This observation enables us to apply results from monitored quantum circuits to the problem of the black hole interior reconstruction. 
\end{enumerate}

\section{Monitored quantum circuit as sequential measurements}\label{sec:sequential}

We begin by formulating monitored quantum circuits in a generic form that can treat various cases on a unified footing. 

Consider a system of $n$ qubits. Initially the system is in a maximally mixed state $\mu \equiv \frac{I}{d}$ where $d=2^n$ and $I$ is an identity operator. A monitored quantum circuit implements a projective measurement of local Pauli operator $P_{j}'$ and then time-evolves by a unitary operator $U_{j}$ for $j=1,\cdots,\tau$. The circuit can be graphically represented as follows:
\begin{align}
 \figbox{1.5}{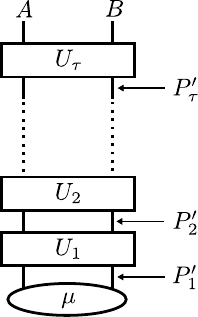}\ .
\end{align}
This setup can characterize various realizations of monitored quantum circuits. For instance, by taking $U_{j}=I$, one can account for the cases where multiple Pauli measurements are performed simultaneously. Also, if one hopes to study the cases where the initial states are product states instead of a maximally mixed state, one may measure all the $n$ qubits with local Pauli operators at the beginning. 

Instead of using local Pauli operators $P_{j}'$ and time-evolution unitary operators $U_{j}$, it is convenient to consider time-evolved Pauli operators:
\begin{align}
P_{j} \equiv (U_{\tau} \cdots U_{j}) P_{j}' (U_{\tau} \cdots U_{j})^{\dagger}.
\end{align}
These time-evolved operators satisfy the following relation:
\begin{align}
P_{\tau}\cdots P_{1} = U_{\tau} P_{\tau}' U_{\tau-1} P_{\tau-1}' \cdots P_{2}'U_{1} P_{1}' ( U_{\tau}\cdots U_{1} )^{\dagger}. 
\end{align}
Note that $( U_{\tau}\cdots U_{1} )^{\dagger}$ act trivially on the maximally mixed state $\mu$. Hence, a monitored circuit can be formulated simply as sequential measurements of time-evolved Pauli operators $P_{j}$ for $j=1,\cdots, \tau$. It is worth mentioning that this formulation can handle the measurement-only circuits~\cite{Ippoliti:2021ab} as well. 

When a monitored circuit time-evolves by Clifford unitary operators, $P_{j}$ are always Pauli operators (since the Clifford unitary operators transform Pauli operators into Pauli operators by its definition). Measurement projection operators are defined by
\begin{align}
\Pi_{j}(m_{j}) \equiv \frac{I+ m_{j}P_{j}}{2} \qquad m_{j} = \pm 1
\end{align}
where $m_{j}=\pm 1$ corresponds to the measurement outcomes. A monitored quantum circuit simply implements the following quantum operation
\begin{align}
\Pi(m) \equiv \Pi_{\tau}(m_{\tau})\cdots \Pi_{1}(m_{1})
\end{align}
where $m$ collectively denotes the measurement outcomes $m = (m_{1},\cdots,m_{\tau})$. Namely, it can be expressed as the following quantum channel:
\begin{align}
\mathcal{Q}\big(\cdot\big) = \sum_{m} \Pi(m)\big( \cdot \big) \Pi^{\dagger}(m).
\end{align}

The probability of measuring $m$ is given by
\begin{align}
\text{Prob}(m) = \big\langle \Pi(m)^{\dagger} \Pi(m) \big\rangle = \frac{1}{d}\Tr\big[ \Pi(m)^{\dagger} \Pi(m) \big]
\end{align}
which can be graphically represented as follows:
\begin{align}
\text{Prob}(m) = \  \figbox{1.5}{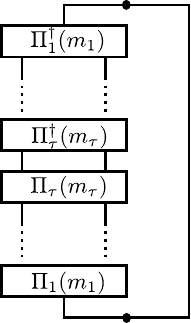} \ 
\end{align}
where each black dot represents a factor of $\frac{1}{\sqrt{d}}$.

In this paper, we are particularly interested in the entanglement structure of the output quantum state of a monitored quantum circuit. To be concrete, let us divide the Hilbert space into two subsystems $A$ and $B$ where $A$ is a smaller subsystem. Here it is convenient to introduce a reference system $R$ and purify the whole system. Then the output state of a monitored circuit is given by
\begin{align}
|\Psi(m)\rangle = \frac{1}{\sqrt{\Prob(m)}} \  \figbox{1.5}{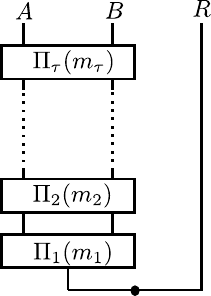} \ . \label{eq:output_monitor}
\end{align}
This expression is valid only when $\Prob(m)\not=0$. In the next several sections, we will develop a theoretical framework that enables us to study and verify the entanglement structure among subsystems $A,B,R$ in monitored Clifford circuits. 

\section{Dual classical error-correction problem}\label{sec:classical_code}

In this section and the next two sections, we discuss the entanglement structure between two subsystems $A$ and $B$. In this section, we will introduce a certain classical error-correction problem that is essential in studying the entanglement structure of a monitored Clifford circuit.

\subsection{Codeword and error vectors}

We begin by introducing certain vectors which record commutation relations among Pauli operators $P_{A}$ supported on the subsystem $A$ and measured Pauli operators $P_{j}$ in the past.  

For Pauli operators $P_{A}\in \Pauli_A$, we assign $\pm 1$ using its commutation relations with respect to $P_{j}$ as follows:
\begin{align}
\C(P_{A})_{j} = \pm 1 \qquad P_{A} P_{j} =  \pm P_{j} P_{A} \qquad (j=1,\cdots,\tau).
\end{align}
We denote them collectively as vectors:
\begin{align}
\C(P_{A}) = \big(  \C(P_{A})_1, \cdots, \C(P_{A})_{\tau}  \big)
\end{align}
and call them \emph{codeword vectors}.

As for $P_{i}$, we assign $\pm 1$ according to its commutation relations with respect to other measured Pauli operators $P_{j}$ as follows:
\begin{equation}
\begin{split}
&\E(P_{i})_{j} = 1 \qquad \qquad \qquad \qquad \qquad \qquad \ \ \ (j > i) \\
&\E(P_{i})_{j} = \pm 1 \qquad P_{i} P_{j} =  \pm P_{j} P_{i} \qquad \qquad (j \leq i).
\end{split}
\end{equation}
Again we denote them collectively as vectors:
\begin{align}
\E(P_{i}) = \big(  \E(P_{j})_1, \cdots, \E(P_{j})_{\tau}  \big)
\end{align}
and call them \emph{error vectors}. Here it is worth emphasizing that, if $i < j$, $\E(P_{i})_{j} = 1$ regardless of the commutation relation between $P_{i}$ and $P_{j}$. In other words, we will look at commutation relations with respect to operators in the past only, and not those in the future. So, the causal orderings of $P_{j}$ are important. 

Let us introduce a few more notations. We will consider the \emph{error vector set} which is generated by component-wise multiplications of $\E(P_{j})$~\footnote{
In this paper, we mainly use ``spin variables'' instead of ``binary variables'' since spin variables are particularly useful in dealing with Pauli operators. For a spin variable $m_{j} =\pm 1 $, we can associate the corresponding binary variables as follows:
\begin{align}
m_{j} =\pm 1  \qquad b(m_{j})\equiv \frac{1-m_{j}}{2} = 0,1.
\end{align}
It is convenient to define ``multiplications'' and ``summations'' for these variables. Namely we have 
\begin{align}
m_{i}\cdot m_{j} \qquad \leftrightarrow \qquad b(m_{i}) + b(m_{j})
\end{align}
where the summation is modulo $2$.
}
:
\begin{align}
\E \equiv \Big\langle \big\{ \E(P_{j}) \big\}_{\forall j} \Big\rangle.
\end{align}
One can also define the following sets of vectors which are generated by acting error vectors $\E(P_{i})$ on a codeword vector $\C(P_{A})$:
\begin{align}
\E^{(P_{A})} \equiv \Big\{ e \cdot \C(P_{A})  : e\in \E \Big\}.
\end{align}
Note $\E^{(I_{A})} = \E$. Finally it will be convenient to introduce the joint set of $\E^{(P_{A})}$:
\begin{align}
\E_{\text{total}} = \bigcup_{P_{A}\in \Pauli_{A}} \E^{(P_{A})}.
\end{align}

\subsection{Classical error-correcting code}\label{sec:code:interpretation}

The above vectors $\C(P_{A})$ and $\E(P_{j})$ can be interpreted as codeword and error vectors in a classical error-correcting code. 

To see this explicitly, assume that the subsystem $A$ consists of $n_{A}$ qubits. There are $4^{n_{A}}$ different Pauli operators on $A$, which can be viewed as $2n_{A}$ bits of classical information~\footnote{For instance, when $n_{A}=1$, we can assign $(1,0)$ and $(0,1)$ to Pauli $X$ and $Z$ operators respectively.}. Let us think of encoding this $2n_{A}$ bits of classical information into $\tau$ physical bits. Here codewords are chosen according to commutation relations between a Pauli operator $P_{A}$ on $A$ and $P_{j}$'s:
\begin{align}
P_{A}\in \Pauli_A \quad \xrightarrow{\text{encode}} \quad \C(P_{A}) = \big( \C(P_{A})_{1},\cdots,  \C(P_{A})_{\tau} \big).
\end{align}
This code attempts to encode $k=2n_{A}$ logical bits into $\tau$ physical bits. In order for this code to be non-trivial, the encoding map $P_{A}\rightarrow \C(P_{A})$ needs to be reversible (\emph{i.e.} $P_{A}$ needs to be encoded into a unique codeword  $\C(P_{A})$ for each $P_{A}$). In other words, $P_{A}$ must have unique commutation relation profiles with respect to $P_{j}$.

Next, we discuss error vectors $\E(P_j)$. Imagine that vectors in $\E$ act as possible errors on codeword vectors. To be concrete, assume that the initial codeword was $\C(P_{A})$ and an error $e\in \E$ occurred. The resulting vector is $e\cdot \C(P_{A})$:
\begin{align}
\C(P_{A}) \quad \xrightarrow{\text{error}}\quad e \cdot  \C(P_{A})  \qquad e\in \E.
\end{align}

In order to recover the initial information, one must be able to reverse the action of error vectors: 
\begin{align}
e \cdot  \C(P_{A})  \quad \xrightarrow{\text{recovery?}} \quad \C(P_{A}).
\end{align}
This will be possible when two codeword vectors are not connected by any error vector. Namely, in order for the initial information $P_{A}$ to be fully recoverable, we must have 
\begin{align}
e\cdot \C(P_{A}) \not= f \cdot \C(Q_{A})\qquad \forall e,f\in \E \qquad (P_{A}\not=Q_{A}). \label{eq:EC0}
\end{align}
Otherwise, two codewords $\C(P_{A})$ and $C(Q_{A})$ cannot be reliably distinguished under the action of error vectors. 

The above error-correction condition for full recovery can be rewritten in several equivalent ways as summarized below. 

\begin{enumerate}[1)]
\item For all pairs of Pauli operators $P_{A}, Q_{A}$ with $P_{A}\not= Q_{A}$, we must have
\begin{align}
e \cdot \C(P_{A}) \not= \C(Q_{A}) \qquad \forall e\in \E.
\end{align}
This follows from Eq.~\eqref{eq:EC0} by noting that $e\cdot f \in \E$ for $e,f \in \E$.
\item For all pairs of Pauli operators $P_{A}, Q_{A}$ with $P_{A}\not= Q_{A}$, we must have
\begin{align} 
\E^{(P_{A})}  \cap \E^{(Q_{A})}  = \emptyset.
\end{align}
Here $\E^{(P_{A})}$ can be interpreted as a set of all the vectors which $\C(P_{A})$ may be transformed into by the action of error vectors. Hence, the joint set $\E_{\text{total}}$ must be divisible into $4^{n_{A}}$ distinct cosets $\E^{(P_{A})}$ labelled by $P_{A}\in \Pauli_A$.
\item All the non-identity Pauli operators $P_{A}(\not=I_{A})$ must satisfy 
\begin{align}
\E^{(P_{A})}  \cap \E^{(I_{A})}  = \emptyset.
\end{align}
This follows from the previous condition 2) by noting that the encoding map is linear:
\begin{align}
\C(P_{A}) \cdot \C(Q_{A}) = \C(P_{A}Q_{A}).
\end{align}
Such a classical code is called a linear code.
\item All the non-identity Pauli operators $P_{A}(\not=I_{A})$ must satisfy
\begin{align}
\C(P_{A}) \not\in \E.
\end{align}
This follows from the previous condition 3). If this is not satisfied, the codeword $\C(P_{A})$ would be indistinguishable from the codeword $\C(I_{A})$ when acted by error vectors (since $\E = \E^{(I_{A})}$).
\end{enumerate}

We will mostly use the condition 4) in order to characterize the recoverability of the dual classical error-correcting code.

\subsection{Entanglement structure from classical error-correction}

By studying the dual classical error-correcting code, one can deduce the entanglement structure of a monitored Clifford circuit. Namely, recoverability of the initial information implies the presence of entanglement between $A$ and $B$ as summarized in the following theorem:

\begin{theorem}\label{theorem:entanglement}
In a monitored Clifford circuit, a subsystem $A$ is maximally entangled with its complement $B$ with $I_{(A,B)}=2n_{A}$ if and only if the initial information in the dual classical error-correcting code is fully recoverable.
\end{theorem}

It is worth emphasizing that the theorem applies to arbitrary realizations of measurement outcomes $m=(m_{1},\cdots, m_{\tau})$.

When the classical error-correction condition is not satisfied, two subsystems $A$ and $B$ are not maximally entangled. In these cases, we can still compute a certain entanglement measure between $A$ and $B$. Here, we will focus on the conditional entropy of $A$ given $B$: 
\begin{align}
S_{A|B} \equiv S_{AB} - S_{B}.
\end{align}
Recall that the conditional entropy is positive in classical systems, but can be negative in quantum systems. Namely, it is useful to note
\begin{align}
S_{A|B} = S_{R} - S_{AR} \geq - S_{A} \label{eq:inequality}
\end{align}
where we used the fact that the output quantum state of the monitored circuit is pure on $ABR$ in the first equality. The second inequality used the positivity of the mutual information $I_{(A,R)}\equiv S_{A}+S_{R}-S_{AR}\geq 0$. The equality is achieved when $A$ and $R$ are not correlated at all with $I_{(A,R)}=0$. The minimal value of the conditional entropy is $-n_{A}$, and it is achieved when $A$ and $B$ are maximally entangled with $I_{(A,B)}=2n_{A}$.

When we interpret the outcome $|\Psi(m)\rangle$ as a quantum channel from $A$ to $B$, the conditional entropy $S_{A|B}$ can be viewed as the coherent quantum information of the quantum channel. So, $S_{A|B}$ characterizes how much quantum information can be transmitted from $A$ to $B$ when viewed as a quantum channel. 

Let us denote the value of the conditional entropy for the measurement result of $m=(m_{1},\cdots, m_{\tau})$ by $S_{A|B}(m)$. We will prove the following theorem.

\begin{theorem}\label{theorem:coherent}
The conditional entropy is given by
\begin{align}
S_{A|B}(m) = - n_{A} + \log_{2}  N_{I_{A}}
\end{align}
where $N_{I_{A}}$ is the number of $P_{A} \in \Pauli_{A}$ such that $\C(P_{A})\in \E^{(I_{A})}$.
\end{theorem}

Note that theorem~\ref{theorem:entanglement} follows from theorem~\ref{theorem:coherent}. 

The proof of this theorem will be presented in appendix~\ref{sec:entropy}. When the classical error-correction condition is satisfied, we have $N_{I_{A}}=1$ and $S_{A|B}(m) = - n_{A}$. Then, from Eq.~\eqref{eq:inequality}, we find that $S_{A}=n_{A}$ and thus, $I_{(A,B)}=S_{A} + S_{B} - S_{AB} = 2n_{A}$. Hence, $A$ and $B$ are maximally entangled. It is worth emphasizing that $S_{A|B}(m)$ does not depend on measurement outcomes $m=(m_{1},\cdots, m_{\tau})$. Here it is useful to observe that $N_{I_{A}}$ can be interpreted as the number of lost classical information since $P_{A}$ becomes indistinguishable from $I_{A}$.

Later we will discuss why the conditional entropy $S_{A|B}$, instead of the mutual information $I_{(A,B)}$, can be computed in the framework of using the dual classical error-correcting code.

\subsection{Examples}\label{sec:code:example}

Since codeword vectors and error vectors play particularly important roles in the entanglement structure of a Clifford monitored circuit, it is worth looking at several examples. 

\subsubsection{Commuting $P_{j}$'s}

Let us begin by looking at the case where $[P_{i},P_{j}]=0$. Assume that $n=3$. Assume that $A$ is the first qubit, and $B$ consists of the second and the third qubits. So, we have 
\begin{equation}
\begin{split}
I_{A} &= I_{1}\otimes I_{2}\otimes I_{3} \\
X_{A} &= X_{1}\otimes I_{2}\otimes I_{3} \\
Y_{A} &= Y_{1}\otimes I_{2}\otimes I_{3} \\
Z_{A} &= Z_{1}\otimes I_{2}\otimes I_{3}.
\end{split}
\end{equation}
Let us choose $P_{1},P_{2},P_{3}$ as follows:
\begin{equation}
\begin{split}
P_{1} &= X_1 \otimes X_2 \otimes I_3 \\
P_{2} &= Z_1 \otimes Z_2 \otimes X_3 \\
P_{3} &= Y_1 \otimes Z_2 \otimes Z_3.
\end{split}
\end{equation}
One can check that $P_{j}$'s commute with each other. 

In this monitored circuit, the system starts from the maximally mixed state $\mu=\frac{I}{2^3}$, and then measurements of $P_{1},P_{2},P_{3}$ are performed sequentially. We are interested in whether the subsystem $A$ is entangled with its complement $B$ or not. 

We can construct the codeword vectors and error vectors as follows:
\begin{center}
\begin{tabular}{cccc}
 & $P_{1}$ & $P_{2}$ & $P_{3}$ \\
\hline
$\C(I_{A})$ & $1$  & $1$  & $1$ \\
$\C(X_{A})$ & $1$  & $-1$ & $-1$ \\
$\C(Y_{A})$ & $-1$ & $-1$ & $1$ \\
$\C(Z_{A})$ & $-1$ & $1$  & $-1$ \\
\hline
$\E(P_{1})$ & $1$  & $1$ & $1$ \\
$\E(P_{2})$ & $1$ & $1$ & $1$ \\
$\E(P_{3})$ & $1$ & $1$  & $1$ 
\end{tabular}
\end{center}
We see that all the error vectors are trivial; $(1,1,1)$. Hence the error vector set is given by
\begin{align}
\mathcal{E} = \big\{ (1,1,1) \big\}.
\end{align}
Also, observe that codeword vectors are unique. Hence, we have 
\begin{equation}
\begin{split}
\mathcal{E}^{(I_{A})} &= \big\{ (1,1,1) \big\} \\
\mathcal{E}^{(X_{A})} &= \big\{ (1,-1,-1) \big\} \\
\mathcal{E}^{(Y_{A})} &= \big\{ (-1,-1,1) \big\} \\
\mathcal{E}^{(Z_{A})} &= \big\{ (-1,1,-1) \big\}
\end{split}
\end{equation}
which do not overlap with each other. We saw that $P_{A}$ is encoded into the codewords $\C(P_{A})$ in a unique manner, and the error from $\E$ cannot connect different codewords. Hence the initial information about $P_{A}$ is recoverable, which implies that $A$ is maximally entangled with $B$. 

As this example suggests, when $[P_{i},P_{j}]=0$ for all $i,j$, $A$ and $B$ are maximally entangled if and only if the codewords $\C(P_{A})$ are unique for different $P_{A}$. This was originally pointed out~\cite{Yoshida:21a} in the context of the Hayden-Preskill recovery problem.

\subsubsection{Non-commuting $P_{j}$'s (recoverable)}

Next, let us choose $P_{1},P_{2},P_{3}$ as follows:
\begin{equation}
\begin{split}
P_{1} &= X_1 \otimes Z_2 \otimes I_3 \\
P_{2} &= Z_1 \otimes I_2 \otimes X_3 \\
P_{3} &= I_1 \otimes X_2 \otimes X_3.
\end{split}
\end{equation}
Codeword vectors and error vectors are given as follows:
\begin{center}
\begin{tabular}{cccc}
 & $P_{1}$ & $P_{2}$ & $P_{3}$ \\
\hline
$\C(I_{A})$ & $1$  & $1$  & $1$ \\
$\C(X_{A})$ & $1$  & $-1$ & $1$ \\
$\C(Y_{A})$ & $-1$ & $-1$ & $1$ \\
$\C(Z_{A})$ & $-1$ & $1$  & $1$ \\
\hline
$\E(P_{1})$ & $1$  & $1$ & $1$ \\
$\E(P_{2})$ & $-1$ & $1$ & $1$ \\
$\E(P_{3})$ & $-1$ & $1$  & $1$ 
\end{tabular}
\end{center}
The error vector set is given by
\begin{align}
\mathcal{E} = \big\{ (1,1,1), (1,-1,-1) \big\}.
\end{align}
We also have
\begin{equation}
\begin{split}
\mathcal{E}^{(I_{A})} &= \big\{ (1,1,1), (1,-1,-1) \big\} \\
\mathcal{E}^{(X_{A})} &= \big\{ (1,-1,1), (1,1,-1) \big\} \\
\mathcal{E}^{(Y_{A})} &= \big\{ (-1,-1,1), (-1,1,-1)\big\} \\
\mathcal{E}^{(Z_{A})} &= \big\{ (-1,1,1), (-1,-1,-1)\big\}
\end{split}
\end{equation}
which do not overlap with each other. Hence, the codewords $\C(P_{A})$ are recoverable under the errors from $\E$. In this case, $A$ is maximally entangled with $B$.

\subsubsection{Non-commuting $P_{j}$'s (not recoverable)}

Let us choose $P_{1},P_{2},P_{3}$ as follows:
\begin{equation}
\begin{split}
P_{1} &= X_1 \otimes I_2 \otimes Z_3 \\
P_{2} &= Z_1 \otimes Z_2 \otimes Z_3 \\
P_{3} &= Y_1 \otimes Z_2 \otimes Z_3.
\end{split}
\end{equation}
We then have
\begin{center}
\begin{tabular}{cccc}
 & $P_{1}$ & $P_{2}$ & $P_{3}$ \\
\hline
$\C(I_{A})$ & $1$  & $1$  & $1$ \\
$\C(X_{A})$ & $1$  & $-1$ & $-1$ \\
$\C(Y_{A})$ & $-1$ & $-1$ & $1$ \\
$\C(Z_{A})$ & $-1$ & $1$  & $-1$ \\
\hline
$\E(P_{1})$ & $1$  & $1$ & $1$ \\
$\E(P_{2})$ & $-1$ & $1$ & $1$ \\
$\E(P_{3})$ & $-1$ & $-1$  & $1$ 
\end{tabular}
\end{center}

The error vector set is given by
\begin{align}
\mathcal{E} = \big\{ (1,1,1), (-1,1,1), (-1,-1,1) , (1,-1,1)  \big\}.
\end{align}
We also have
\begin{equation}
\begin{split}
\mathcal{E}^{(I_{A})} &= \big\{  (1,1,1), (-1,1,1), (-1,-1,1) , (1,-1,1)  \big\} \\
\mathcal{E}^{(X_{A})} &= \big\{  (1,-1,-1), (-1,-1,-1), (-1,1,-1) , (1,1,-1)  \big\} \\
\mathcal{E}^{(Y_{A})} &= \big\{ (-1,-1,1), (1,-1,1), (1,1,1) , (-1,1,1) \big\} \\
\mathcal{E}^{(Z_{A})} &= \big\{  (-1,1,-1), (1,1,-1), (1,-1,-1) , (-1,-1,-1) \big\}
\end{split}
\end{equation}
which are not distinct. Hence, the codewords $\C(P_{A})$ are not recoverable under the errors from $\E$. In this case, $A$ is not maximally entangled with $B$. Namely, we will have $S_{A|B}=0$.

\section{Entanglement distillation between two subsystems}\label{sec:distillation}

In this section, we will describe the entanglement distillation algorithm and compute its output.

\subsection{Perfect distillation}

To build some intuition, we begin by discussing the cases where $A$ and $B$ are maximally entangled (\emph{i.e.} the classical error-correction condition is satisfied). 

The distillation algorithm proceeds in a way similar to algorithms from~\cite{Yoshida:2017aa, Yoshida:21a, Yoshida:2021aa}. The overall procedure is graphically summarized as follows:
\begin{align}
 \figbox{1.5}{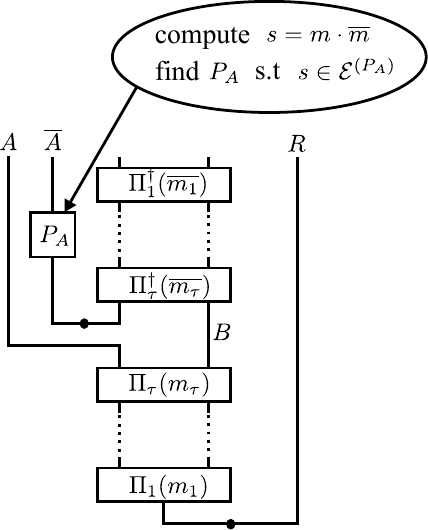}. \
\end{align}
Given the outcome of the monitored circuit $|\Psi(m)\rangle$, we keep qubits on $A$ aside and add EPR pairs on $A\overline{A}$. Then we performs projective measurements of $P_{\tau}^{\dagger},\cdots, P_{1}^{\dagger}$ whose measurement outcomes are denoted by $\overline{m}$. This process can be written as $\Pi(\overline{m})^{\dagger}$. Finally, some appropriate feedback operation is applied on $\overline{A}$ based on the measurement results $m$ and $\overline{m}$. 

Let us discuss how to construct an appropriate feedback operator. It is convenient to define a sum vector $s$ via component-wise multiplications: 
\begin{align}
s \equiv m \cdot \overline{m}.
\end{align}
We will prove that the measurement of $m$ and $\overline{m}$ may occur only when $s=m\cdot \overline{m} \in \E_{\text{total}}$ where $\E_{\text{total}} = \bigcup_{P_{A}}\E^{(P_{A})}$. In fact, we can compute the probability of measuring $s$ explicitly. Let us denote the probability of measuring $m$ and $\overline{m}$ by $\Prob(m,\overline{m})$. 
It is convenient to define the summation of probabilities over $m$ as follows:
\begin{align}
\Sum(s) \equiv \sum_{m}\Prob(m, m\cdot s )
\end{align}
which corresponds to the total probability of measuring $s$. We will prove the following lemma.

\begin{lemma}\label{lemma:sum}
The probability of measuring $s$ is given by  
\begin{equation}
\begin{split}
\Sum(s) \equiv \sum_{m}\Prob(m, m\cdot s ) &= \frac{1}{d_{\E_{\text{total}}}} \qquad s\in \E_{\text{total}} \\ 
&= 0 \qquad \ \ \quad \ s\not\in \E_{\text{total}}
\end{split}
\end{equation}
where $d_{\E_{\text{total}}}$ is the number of elements in the joint set $\E_{\text{total}}$. 
\end{lemma}

So, measurement of $s$ with $s\not\in \E_{\text{total}}$ will never occur. The proof of this lemma will be presented in appendix~\ref{sec:probability}.

Now we discuss how to construct a feedback operator. One immediate corollary of lemma~\ref{lemma:sum} is that one can always find $P_{A}\in \Pauli_A$ such that
\begin{align}
s \in \E^{(P_{A})}.
\end{align}
It turns out that the necessary feedback operation is to simply implement $P_{A}$ on $\overline{A}$. In general, there can be multiple $P_{A}$ which satisfy $s \in \E^{(P_{A})} $. But, when the classical error-correction condition is satisfied, then one can always find a unique $P_{A}$ satisfying $s\in \E^{(P_{A})}$. Hence, the task of finding $P_{A}$ can be interpreted as decoding of the initial classical information $P_{A}$ from a bit string $s$ in the dual classical code.

One comment follows. When the measurement result satisfies $s \in \E = \E^{(I_{A})}$, there is no need of applying a feedback operation. If the classical error-correction condition is satisfied, this occurs with the following probability
\begin{align}
\frac{\sum_{s \in \E^{(I_{A})}}\Sum(s)}{\sum_{s\in \E_{\text{total}}}\Sum(s)} = \frac{1}{d_{A}^2}.
\end{align}
This probability matches with the successful post-selection decoding probability for the Hayden-Preskill decoding algorithm~\cite{Yoshida:2017aa}.

Here we summarize the distillation algorithm.

\begin{enumerate}
\item Given the outcome of the monitored circuit $|\Psi(m)\rangle$, keep qubits on $A$ aside and insert ancilla EPR pairs on $\overline{A}$ and $A$. 
\item Perform measurements of $P_{\tau}^{\dagger},\cdots,P_{1}^{\dagger}$ by applying $\Pi^{\dagger}(\overline{m})$.
\item Compute $s=m\cdot \overline{m}$ and find $P_{A}\in \Pauli_A$ such that $s\in \E^{(P_{A})}$. 
\item Apply $P_{A}$ on $\overline{A}$. Perfect EPR pairs will be distilled on $A$ and $\overline{A}$ if the classical error-correction condition is satisfied. 
\end{enumerate}

\subsection{Imperfect distillation}

If the recoverability condition is not satisfied, the outcome of the distillation algorithm will prepare imperfect EPR pairs. Here we will explicitly compute the output state, averaged over all the possible measurement results $m,\overline{m}$.

Let us denote the output of the aforementioned distillation algorithm by $\sigma_{A\overline{A}}(m,
\overline{m})$. We are particularly interested in its statistical average defined by
\begin{align}
\mathbb{E}\big( \sigma_{A\overline{A}} \big) \equiv \sum_{m,\overline{m}} \Prob(m,\overline{m})\sigma_{A\overline{A}}(m,\overline{m}).
\end{align}
The averaged output quantum state can be computed explicitly as follows:

\begin{lemma}\label{lemma:output}
The output of the aforementioned distillation algorithm for a monitored Clifford circuit is 
\begin{align}
\mathbb{E}\big(\sigma_{A\overline{A}}\big) = \frac{1}{N_{I_{A}}} \sum_{P_{A}:\C(P_{A})= \E^{(I_{A})} } | P_{A}\rangle \langle P_{A} |
\end{align}
where $N_{I_{A}}$ is the number of $P_{A}$ such that $\C(P_{A})= \E^{(I_{A})}$. 
\end{lemma}

Here, $|P_{A}\rangle$ represents the Choi-Jamio\l{}kowski state of $P_{A}$, namely
\begin{align}
|P_{A}\rangle \equiv (P_{A}\otimes I_{\overline{A}}) |\text{EPR}\rangle_{A\overline{A}}.
\end{align}
Note that $|P_{A}\rangle$'s form a complete orthonormal basis for $A$ and $\overline{A}$. The proof of this lemma will be presented in appendix~\ref{sec:output}.

The statistical average $\mathbb{E}(\sigma_{A\overline{A}})$ can capture quantum entanglement between $A$ and $B$ even though it is averaged over all the possible realizations of $m$ and $\overline{m}$. Let us compute the conditional entropy $S_{A|\overline{A}}$ for $\mathbb{E}(\sigma_{A\overline{A}})$:
\begin{align}
\text{$S_{A|\overline{A}}$ of $\mathbb{E}(\sigma_{A\overline{A}})$} = \log N_{I_A} - n_{A}
\end{align}
which matches with the value from theorem~\ref{theorem:coherent}:
\begin{align}
S_{A|B}(m) =  \log N_{I_A} - n_{A}.
\end{align}

\subsection{Examples}

It will be useful to look at concrete examples in order to gain some intuitions. For simplicity of discussion, we will focus on systems with two qubits ($n=2$) with $n_{A}=n_{B}=1$.

\subsubsection{No measurement}

Let us begin with the most trivial case. If no measurement is performed at all, the output is
\begin{align}
\mu_{A} \otimes \mu_{B}
\end{align}
where $\mu_{A}$ and $\mu_{B}$ are maximally mixed states on $A$ and $B$ respectively. The conditional entropy is
\begin{align}
S_{A|B} = 1. 
\end{align}

\subsubsection{Single-qubit measurement}

Assume that there was only a single measurement with $\tau=1$, and it was with $P_{1}=Z_{A}$. In this case, the output of the monitored circuit is 
\begin{equation}
\begin{split}
&|0\rangle\langle 0|_{A}\otimes \mu_{B}  \qquad m_{1}=1 \\
&|1\rangle\langle 1|_{A}\otimes \mu_{B}  \qquad m_{1}=-1.
\end{split}
\end{equation}
One can see that $A$ and $B$ have no correlation at all. For both cases, we have
\begin{align}
S_{A|B}(m_{1}) = 0 \qquad m_{1}=\pm 1.
\end{align}

The output for the entanglement distillation algorithm is 
\begin{equation}
\begin{split}
&|0\rangle\langle 0|_{A}\otimes |0\rangle\langle 0|_{\overline{A}}  \qquad m_{1}=1 \\
&|1\rangle\langle 1|_{A}\otimes |1\rangle\langle 1|_{\overline{A}}  \qquad m_{1}=-1.
\end{split}
\end{equation}
Hence its average over $m_{1}$ is given by
\begin{align}
\mathbb{E}\big(\rho_{A|\overline{A}}\big) = \frac{1}{2} \big( |00\rangle \langle 00 | + |11\rangle \langle 11 | \big) 
\end{align}
which possesses classical correlation. Note that this classical correlation was generated by taking an average over $m_{1}$. Finally we can see that the conditional entropy is given by
\begin{align}
\text{$S_{A|\overline{A}}$ of $\mathbb{E}\big(\sigma_{A|\overline{A}}\big)$ } = 0.
\end{align}

It is worth computing the mutual information. For the output of a monitored circuit, we have
\begin{align}
I_{(A,B)} (m_{1}) = 0 \qquad m_{1} = \pm 1
\end{align}
where $I_{(A,B)}\equiv S_{A} + S_{B} - S_{AB}$. On the other hand, for the averaged output of the distillation algorithm, we have 
\begin{align}
\text{$I_{(A,\overline{A})}$ of $\mathbb{E}(\sigma_{A|\overline{A}})$} = 1
\end{align}
due to the classical correlation. Hence, the values of the mutual information do not match.

\subsubsection{Two-qubit measurement}

Next, assume that there was only a single measurement with $\tau=1$, and it was with $P_{1}=Z_{A} \otimes Z_{B}$. In this case, the output of the monitored circuit is 
\begin{equation}
\begin{split}
&\frac{1}{2} \big( |00\rangle  \langle 00| +  |11\rangle \langle 11| \big) \qquad m_{1}=1 \\
& \frac{1}{2} \big( |01\rangle\langle 01| + |10\rangle \langle 10| \big)  \qquad m_{1}=-1.
\end{split}
\end{equation}
One can see that $A$ and $B$ share classical correlation which was induced by measurement of $Z_{A} \otimes Z_{B}$. We find
\begin{align}
S_{A|B}(m_{1}) = 0 \qquad m_{1}=\pm 1.
\end{align}

The output for the entanglement distillation algorithm is 
\begin{equation}
\begin{split}
\frac{1}{2} \big( |00\rangle \langle 00 | + |11\rangle \langle 11 | \big) \qquad m_{1} = \pm 1.
\end{split}
\end{equation}
Note that classical correlation is present even without taking an average over $m_{1}$. We see that the conditional entropy is given by
\begin{align}
\text{$S_{A|\overline{A}}$ of $\mathbb{E}\big(\sigma_{A|\overline{A}}\big)$ } = 0.
\end{align}

As for the mutual information, we have 
\begin{align}
I_{(A,B)} (m_{1}) = 1 \qquad m_{1} = \pm 1
\end{align}
and
\begin{align}
\text{$I_{(A,\overline{A})}$ of $\mathbb{E}(\sigma_{A|\overline{A}})$} = 1.
\end{align}
Hence, the values of the mutual information match. 

One important lesson from this and previous examples is that the monitored quantum circuits generate different output states in two examples, but the averaged output $\mathbb{E}(\sigma_{A\overline{A}})$ from the distillation algorithm is the same for both cases. This is because $Z_{A}\otimes I_{B}$ and $Z_{A}\otimes Z_{B}$ have the same patterns of commutation relations with $P_{A}$. This is also the reason why the conditional entropy, instead of the mutual information, is computable from the dual classical code.

\subsubsection{Two-qubit commuting measurement}

Next, assume that we perform two commuting measurements with $P_{1}=X_{A}\otimes X_{B}$ and $P_{2}=Z_{A}\otimes Z_{B}$. In this case, the output of the monitored circuit is 
\begin{equation}
\begin{split}
|I_{A}\rangle, |X_{A}\rangle, |Y_{A}\rangle, |Z_{A}\rangle \qquad (m_{1},m_{2}) = (1,1), (1,-1), (-1,-1), (-1,1)
\end{split}
\end{equation}
where $|P_{A}\rangle \equiv(P_{A}\otimes I_{B})|\text{EPR}\rangle_{AB}$. All the possible output states are maximally entangled, and we have
\begin{align}
S_{A|B}(m_{1},m_{2}) = - 1 \qquad m_{1},m_{2}=\pm 1.
\end{align}

The outputs for the entanglement distillation algorithm are 
\begin{equation}
\begin{split}
|I_{A}\rangle = |\text{EPR}\rangle_{AB} \qquad m_{1},m_{2}=\pm 1
\end{split}
\end{equation}
and the conditional entropy is given by
\begin{align}
\text{$S_{A|\overline{A}}$ of $\mathbb{E}\big(\sigma_{A|\overline{A}}\big)$ } = -1.
\end{align}

In this case, the codeword vectors are 
\begin{equation}
\begin{split}
\C(I_{A}) = (1,1)\quad \C(X_{A}) = (1,-1) \quad \C(Y_{A}) = (-1,-1)\quad \C(Z_{A}) = (-1,1).
\end{split}
\end{equation}
Also, the error vector set is trivial $\E=\{(1,1)\}$ because $P_{1}$ and $P_{2}$ commute. Hence, the classical error-correction condition is satisfied.

\subsubsection{Two-qubit non-commuting measurement}

Finally, assume that we perform two on-commuting measurements with $P_{1}=X_{A}\otimes Z_{B}$ and $P_{2}=Z_{A}\otimes Z_{B}$. In this case, the output of the monitored circuit is 
\begin{equation}
\begin{split}
&\frac{1}{2} \big( |00\rangle  \langle 00| +  |11\rangle \langle 11| \big) \qquad m_{1}=\pm1, \quad m_{2}=1 \\
& \frac{1}{2} \big( |01\rangle\langle 01| + |10\rangle \langle 10| \big)  \qquad m_{1}=\pm1, \quad m_{2}=-1.
\end{split}
\end{equation}
Note that the measurement result $m_{1}$ does not affect the output state. The outputs for the entanglement distillation algorithm is 
\begin{equation}
\begin{split}
\frac{1}{2} \big( |00\rangle \langle 00 | + |11\rangle \langle 11 | \big) \qquad m_{1}=\pm1, \quad m_{2} = \pm 1.
\end{split}
\end{equation}
We see that the two subsystems share classical correlations only.

In this case, the codeword vectors are 
\begin{equation}
\begin{split}
\C(I_{A}) = (1,1)\quad \C(X_{A}) = (1,-1) \quad \C(Y_{A}) = (-1,-1)\quad \C(Z_{A}) = (-1,1).
\end{split}
\end{equation}
The error vector set is trivial $\E=\{(1,1), (-1,1)\}$ because $P_{1}$ and $P_{2}$ anti-commute. Hence, the classical error-correction condition is not satisfied. We have 
\begin{equation}
\begin{split}
\E^{(I_{A})} = \E^{(Z_{A})} = \{(1,1), (-1,1)\} \qquad  \E^{(X_{A})} = \E^{(Y_{A})} = \{(1,-1), (-1,-1)\}
\end{split}
\end{equation}
which suggests that the output of the distillation algorithm is
\begin{align}
\frac{1}{2} \big( |I_{A}\rangle \langle I_{A} | + |Z_{A}\rangle \langle Z_{A} | \big) = \frac{1}{2} \big( |00\rangle \langle 00 | + |11\rangle \langle 11 | \big).
\end{align}

\subsection{Distillation algorithm for the Gullans-Huse proposal}

Let us apply the aforementioned distillation algorithm to the proposal by Gullans and Huse which entangles the system to a single reference qubit~\cite{Gullans:2020ab}. 

Consider an EPR pair on $R\overline{R}$ where each of $R$ and $\overline{R}$ consists of a single qubit. Here we think of encoding $\overline{R}$ into $n$ qubits by some Clifford isometry and use it as an initial state of the monitored Clifford circuit. One can represent the output wavefunction as follows:
\begin{align}
 \figbox{1.5}{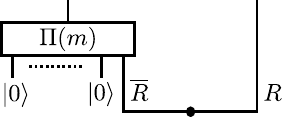}  \label{eq:GH}
\end{align}
by adding $n-1$ ancilla qubits prepares in $|0\rangle^{\otimes n-1}$. Here the encoding circuit is absorbed into the definition of measured Pauli operators $P_{j}$. The key idea of the Gullans-Huse proposal is that the entanglement between the system and the reference will survive for long time when the monitored quantum circuit is in the volume-law phase. Hence, the distillability of an EPR pair serves as an order parameter to detect the dynamical entanglement phase transition. Our goal is to construct an algorithm to distill an EPR pair from the system and the reference in this setup. 

The aforementioned algorithm was designed to distill the entanglement within the system. In order to apply it to the Gullans-Huse proposal, we will view the whole of $n+1$ qubits (including the reference $R$) as the ``system'' of the monitored Clifford circuit. Namely, we imagine that the system was initially in the maximally mixed state, and then we performed measurements of $Z_{1}, \cdots, Z_{n-1}$ and Bell measurements $X_{n}\otimes X_{n+1}$ and $Z_{n}\otimes Z_{n+1}$:
\begin{align}
 \figbox{1.5}{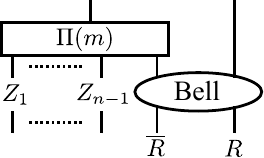} \ . 
\end{align}
Eq.~\eqref{eq:GH} can be obtained by postselecting the measurement outcomes to $Z_{j}=+1$ and $X_{n}\otimes X_{n+1} = Z_{n}\otimes Z_{n+1} = +1$. In this interpretation, we need to extend codeword vectors and error vectors as follows
\begin{align}
\C(P_{R}) = ( \text{Bell}, \text{$Z_{j}$'s}, \text{$P_{j}$'s} )
\end{align}
so that commutation relations with Bell operators and $Z_{j}$'s at the beginning are taken into account. The distillation algorithm is given by
\begin{align}
 \figbox{1.5}{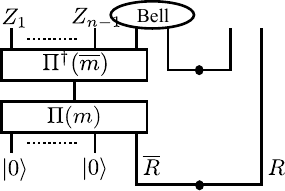}  
\end{align}
where measurements of $Z_{j}$'s and Bell operators are performed at the very end of the algorithm. The Bell measurements at the very end can be omitted, leading to the following simplified algorithm:
\begin{align}
 \figbox{1.5}{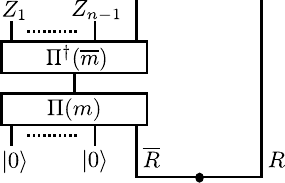}  
\end{align}
This simplification has an effect of restricting the sum vector $s$ to be
$s = ( +1, \text{$Z_{j}$'s}, \text{$P_{j}$'s} )$.
By decoding this sum vector, one can obtain an appropriate feedback Pauli operator $P_{R}$ to distill an EPR pair on $R\overline{R}$ (if the system and the reference remain entangled).

\section{Operator growth in spacetime}\label{sec:scrambling}

Previous works have noted that the scrambling dynamics in a monitored quantum circuit underpins the emergence of the volume-law entanglement. While this speculation would have profound implications, concrete arguments establishing this connection have not been presented. Indeed, entanglement creation in unitary quantum circuits is a process of thermalization that has no direct relevance to quantum information scrambling~\footnote{This confusion can be found in earlier works on the fast scrambling conjecture, see~\cite{Sekino08} for instance. Recent studies have found that entanglement creation may not occur in the scrambling time scale~\cite{Shor18}. A related observation can be found in~\cite{Hosur:2015ylk} as well.}. Namely, unlike thermalization which concerns the time-evolution of a quantum state, quantum information scrambling stems from the growth of local operators which can be quantitatively measured by using out-of-time order correlation (OTOC) functions~\cite{Hosur:2015ylk, Kitaev_unpublished, Roberts:2015aa}. 

\begin{figure}
\centering
\includegraphics[width=0.35\textwidth]{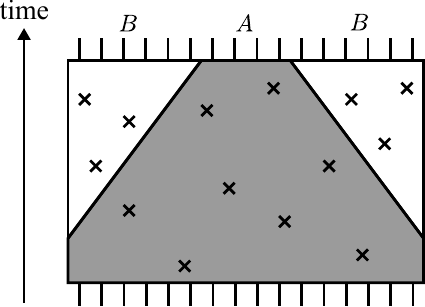}
\caption{Operator growth and entanglement creation in a monitored quantum circuit. Cross marks represent measured Pauli operators. A local Pauli operator $P_{A}$ on $A$ is encoded into a codeword vector by overlapping with measured Pauli operators in the past which lie inside the shaded region of the spacetime. 
}
\label{fig-scrambling}
\end{figure}

Our characterization establishes a direct and concrete relation between entanglement in monitored quantum circuits and the operator growth. A central object in our analysis was the codeword vector $\C(P_{A})$ which can be understood as OTOC functions:
\begin{align}
\C(P_{A})_{j} = \langle P_{A}P_{j}P_{A}^{\dagger}P_{j}^{\dagger} \rangle.
\end{align}
For the subsystem $A$ to be entangled with $B$, the underlying dynamic (the unitary part of the monitored circuit) needs to be scrambling. Namely, $P_{A}$ should evolve back and overlap non-trivially with measured Pauli operators $P_{j}$ in the past so that the codeword vector $\C(P_{A})$ is non-trivial and resilient against errors (see Fig.~\ref{fig-scrambling}). A crucial point is that, without local $P_{j}$ measurements, the subsystem $A$ would be entangled with the reference $R$. Local projective measurements $P_{j}$ decouple $A$ from the reference $R$, and instead make $A$ entangled with its complement $B$. 

Here, it is important to emphasize that entanglement creation in the volume-law phase is a result of a subtle competition between the decoupling phenomena and accumulations of error vectors in $\mathcal{E}$. Namely, while overlapping with operators in the past is crucial for robust codeword vectors $\mathcal{C}(P_{A})$, too many projective measurements will make the error vector set $\mathcal{E}$ rather dense and bring the system to the area-law phase. Also, our analyses in this paper so far primarily focus on Clifford circuits. It is worth noting, however, that the decoupling phenomena, which disentangles $A$ from $R$, is a generic feature of scrambling systems, and is not restricted to Clifford circuits~\cite{Beni18}. We expect that the space-time pattern of the operator growth will be an interesting subject of study, and hope to further establish the connection between scrambling dynamics and entanglement creation in monitored quantum circuits beyond Clifford in a future work. 

\section{Coding properties of monitored Clifford circuit}\label{sec:reference}

So far, we have studied the entanglement between two complementary subsystems without involving the reference system. In this section and the next, we turn our attentions to the entanglement structure of monitored Clifford circuits with the reference system $R$. In this section, we study the coding properties of monitored Clifford circuits. Some additional results are presented in appendix~\ref{app:code} as well.

\subsection{System-Reference entanglement}

Our framework of using a dual classical error-correcting code allows us to study the entanglement structure among subsystems $A,B$ as well as the reference system $R$. 

Recall that we used Pauli operators $P_{A}\in \Pauli_A$ on a subsystem $A$ as initial information of a classical error-correcting code and derived the conditional entropy $S_{A|B}$:
\begin{align}
S_{A|B} = S_{AB} - S_{B} = \log N_{I_{A}} - n_{A} \qquad N_{I_{A}} : \text{number of $P_{A}$ s.t. $\C(P_{A})\in \E $.} \label{eq:A}
\end{align}
One can repeat a similar analysis by choosing $P_{B}\in \Pauli_B$ on a subsystem $B$ as initial information:
\begin{align}
S_{A|B} = S_{AB} - S_{A} = \log N_{I_{B}} - n_{B} \qquad N_{I_{B}} : \text{number of $P_{B}$ s.t. $\C(P_{B})\in \E $.} \label{eq:B}
\end{align}
One can also use all the Pauli operators $P$ supported on $AB$ and treat them as initial information. This leads to 
\begin{align}
S_{AB|\emptyset} = S_{AB} = \log N_{I} - n \qquad N_{I} : \text{number of $P$ s.t. $\C(P)\in \E $.} \label{eq:AB}
\end{align}
Here we interpreted $AB$ as a system of interest so that $AB$'s complement is an empty set $\emptyset$.

These three equations Eq.~\eqref{eq:A}~\eqref{eq:B}~\eqref{eq:AB} are sufficient to specify values of entanglement entropies for all the possible subsystems, namely ($S_{A}$, $S_{B}$, $S_{R}$, $S_{AB}$, $S_{BR}$, $S_{AR}$). Here we compute a few interesting entanglement measures. Let us begin with the mutual information $I_{(A,B)}$:
\begin{align}
I_{(A,B)} = \log \frac{N_{I}}{N_{I_{A}}N_{I_{B}}}
\end{align}
which can be expressed in terms of the numbers of Pauli operators such that $\C(P)\in \E$. 
Namely there may exist a Pauli operator $P= P_{A}\otimes P_{B}$ such that $\C(P_{A}), \C(P_{B}) \not\in \E $, but $\C(P) \in \E $. The above equation suggests that $I_{(A,B)}$ is related to the number of such Pauli operators which are non-local with respect to the bipartition into $A$ and $B$.

Next, let us compute the conditional entropy $S_{AB|R}$:
\begin{align}
S_{AB|R} = n - \log N_{I}
\end{align}
where we used $S_{ABR}=0$. One also finds $I_{(AB,R)} = 2 (\log N_{I}-n)$. Observe that $N_{I}$ is related to the amount of lost information in the dual classical error-correcting code since a Pauli operator $P$ is indistinguishable from an identity operator $I$. It is interesting to note that the entanglement between the system $A$ and the reference $R$ results from the loss of initial information in the dual classical error-correcting code. Intuitively, this result suggests that the lost information, which was not detected by $P_{j}$'s, will flow to the reference $R$. 

As is evident from discussions so far, Pauli operators satisfying $P$ with $\C(P)\in \E$, which is indistinguishable from $I$, play important roles in studying the entanglement structure of a monitored Clifford circuit. We shall call them \emph{null operators} of a dual classical error-correcting code. For later discussions, it will be convenient to define the following three sets of null operators:
\begin{equation}
\begin{split}
\mathcal{L} &\equiv \{ P \in \Pauli : \C(P) \in \E  \} \\
\mathcal{L}_{A} &\equiv \{ P_{A} \in \Pauli_{A} : \C(P_{A}) \in \E  \} \\
\mathcal{L}_{B} &\equiv \{ P_{B} \in \Pauli_{B} : \C(P_{B}) \in \E  \}.
\end{split}
\end{equation}
Note that these sets are actually groups~\footnote{Strictly speaking, a complex phase $iI$ should be included to define a group of Pauli operators. We ignore this subtlety since it is not essential in our treatment. For careful analyses, see~\cite{Bravyi09, Beni10} for instance.}. Later, we will show that these null operators serve as logical operators (including trivial stabilizer operators) when the monitored Clifford circuit is viewed as a quantum error-correcting code. 

\subsection{Stabilizer group}

In this subsection and the next, we will construct stabilizer and logical operators of a monitored Clifford circuit. We begin by constructing the stabilizer group $\mathcal{S}$. The construction proceeds recursively. Here we denote the stabilizer group constructed for $P_{t},\cdots, P_{1}$ by $\Stab^{(t)}$. We start with 
\begin{align}
\Stab^{(1)} \equiv \big\langle P_{1} \big\rangle
\end{align}
and then recursively define
\begin{align}
\Stab^{(\tau)} \equiv \Big\langle P_{\tau}, 
\big\{ P\in \Stab^{(\tau-1)} : [P,P_{\tau}]=0 \big\} \Big\rangle. 
\label{eq:stab_recursive}
\end{align}
It is worth emphasizing that this is different from simply taking the center of the group $\langle \{P_{j}\} \rangle$.

To gain some insight on this construction, let us make the following observations. If $P_{\tau}$ commutes with all the Pauli operators in $\Stab^{(\tau-1)}$, we have 
\begin{align}
\Stab^{(\tau)} = \big\langle P_{\tau}, \Stab^{(\tau-1)} \big\rangle.
\end{align}
On the other hand, if there exists $R \in \Stab^{(\tau-1)} $ such that $\{R,P_{\tau}\}=0$, $R$ is removed and then $P_{\tau}$ will be added to the stabilizer group. In other words, all the Pauli operators which do not commute with $P_{\tau}$ are removed, and instead, $P_{\tau}$ is added to the group. It is useful to note 
\begin{align}
P_{\tau} \in \Stab^{(\tau)} \qquad \text{and} \qquad [Q, P_{\tau}] = 0 \quad \forall Q \in \Stab^{(\tau)}.
\end{align}
So, the last Pauli operator $P_{\tau}$ always enter in the latest stabilizer group $\Stab^{(\tau)}$.

Let us briefly discuss the physical implication of the recursive construction of $\Stab^{(\tau)}$. Observe that the number of operators in $\Stab^{(\tau)}$ increases if and only if $P_{\tau}$ commutes with all the operators in $\Stab^{(\tau-1)}$, and is not included in $\Stab^{(\tau-1)}$. In general, this is not very likely to occur when $\Stab^{(\tau-1)}$ is already large. Namely, if $\dim \Stab^{(\tau-1)} = 2^{n_{S}}$ and $P_{\tau}$ is randomly chosen, the increase will occur only with probability $\approx \frac{1}{2^{n_{S}}}$. Here it is natural to expect that $P_{\tau}$ is a high-weight pseudorandom Pauli operator when the underlying dynamics is scrambling. Hence, the size of $\Stab^{(\tau)}$ will not increase easily once the circuit reaches the entanglement equilibrium. This mechanism is crucial in an exponential memory time in the volume-law phase of monitored quantum circuits.

The constructed group $\mathcal{S}\equiv\Stab^{(\tau)}$ plays the role of the stabilizer group. Given the aforementioned construction, the following lemma can be proven immediately. 

\begin{lemma}\label{lemma:stabilizer_condition}
Let $P \in \mathcal{S}$ be a Pauli operator in the stabilizer group $\mathcal{S}$ of the monitored Clifford circuit. Then we have 
\begin{align}
P |\Psi(m)\rangle = \pm |\Psi(m)\rangle
\end{align} 
where the eigenvalue $\pm1$ depends on the measurement outcome $m$.
\end{lemma}

In a conventional stabilizer code, stabilizer generators $S_{j}$ are chosen so that codeword states are supported on a subspace satisfying $S_{j}|\psi\rangle = + |\psi\rangle$. In a monitored Clifford circuit, the signs of eigenvalues with respect to stabilizer generators depend on the measurement outcome $m$. Given the values of $m$, one can define $\mathcal{S}(m)$ so that $|\Psi(m)\rangle$ is supported on the $+1$ eigenstate space of $\mathcal{S}(m)$ via appropriate relabelling $S_{j} \rightarrow \pm S_{j}$.

\subsection{Logical operators}

Next, we present the construction of a group of null operators
\begin{equation}
\mathcal{L} \equiv \Big\langle P \in \Pauli : \C(P) \in \E \Big \rangle\label{eq:Logic}
\end{equation}
and show that it is the logical operator group. 

Again, the construction proceeds recursively. Here we denote the logical operator group constructed for $P_{t},\cdots, P_{1}$ by $\Logic^{(t)}$. We start with 
\begin{align}
\Logic^{(1)} = \Comm (P_{1}) \equiv \big\langle P \in \Pauli : [P, P_{1}]=0 \big\rangle
\end{align}
where $\Comm$ represents the commutant. Here $\Comm (P_{1})$ contains $2^{2n-1}$ Pauli operators. We then recursively define
\begin{align}
\Logic^{(\tau)} \equiv \Big\langle P_{\tau}, 
\big\{ P\in \Logic^{(\tau-1)} : [P,P_{\tau}]=0 \big\} \Big\rangle. \label{eq:logic_recursive}
\end{align}
In other words, we only keep Pauli operators which commute with $P_{\tau}$ and add $P_{\tau}$ instead. Observe that the stabilizer group $\mathcal{S}^{(\tau)}$ and $\mathcal{L}^{(\tau)}$ are constructed recursively in the same matter in Eq.~\eqref{eq:stab_recursive} and Eq.~\eqref{eq:logic_recursive} except that the initial sets are chosen differently, namely $\Logic^{(1)}=\Comm (\Stab^{(1)})$.

The following lemma will be proven in appendix~\ref{sec:system-proof}.

\begin{lemma}\label{lemma:logical}
We have 
\begin{equation}
\mathcal{L} = \Logic^{(\tau)}
\end{equation}
where $\Logic^{(\tau)}$ is defined recursively via Eq.~\eqref{eq:logic_recursive}. Namely, 
\begin{equation}
\C(P) \in \E \qquad \text{\emph{iff}}\qquad P \in \Logic^{(\tau)}.
\end{equation}
\end{lemma}

Given the aforementioned construction of $\mathcal{L}$, it is immediate to prove that the logical operator group $\mathcal{L}$ is nothing but the commutant of the stabilizer group $\mathcal{S}$.

\begin{corollary}
The logical operator group $\mathcal{L}$ is the commutant of the stabilizer group $\mathcal{S}$:
\begin{align}
\mathcal{L} = \Comm (\mathcal{S}) = \big\{ P\in \Pauli : [P,Q]=0 \ \forall Q\in \mathcal{S} \big\}.
\end{align}
\end{corollary}

Hence, null operators in $\mathcal{L}$ play the role of logical operators when acting on the output wavefunction of a monitored Clifford circuit. In the next section, we will see this more clearly by constructing the Choi-Jamio\l{}kowski state. Note that $\mathcal{L}$ contains trivial logical operators (\emph{i.e.} stabilizer operators) since $\mathcal{S}\subseteq \mathcal{L}$.

Here it is useful to recall that one can choose independent generators of $\mathcal{L}$ as follows~\cite{Beni10}:
\begin{align}
\mathcal{L} = \left\langle\begin{bmatrix}
\overline{Z_{1}} & \cdots & \overline{Z_{k}} & \overline{Z_{k+1}} & \cdots & \overline{Z_{n}} \\
\overline{X_{1}} & \cdots & \overline{X_{k}} & & &
\end{bmatrix}
\right \rangle
\end{align}
where operators commute with each other except for those in the same column. Namely, there always exist some Clifford unitary $U$ which convert above Pauli operators into local ones via 
\begin{align}
\overline{X_{j}} = UX_{j}U^{\dagger} \qquad \overline{Z_{j}} = UZ_{j}U^{\dagger}
\end{align}
up to possible $\pm 1$ signs. Here, the stabilizer group $\mathcal{S}$ is the center of $\mathcal{L}$:
\begin{align}
\mathcal{S} = \Big\langle\overline{Z_{1}}, \cdots, \overline{Z_{k}}  \Big\rangle
\end{align}
since $\mathcal{S} = \Comm(\mathcal{L})$. Here it is useful to note that the double commutant theorem holds for $\mathcal{L}, \mathcal{S}$, namely $\Comm(\Comm(\mathcal{S}))=\mathcal{S}$.

The number of elements in $\Logic^{(\tau)}$ decreases only when $P_{\tau}$ satisfies $[P_{\tau},\Stab^{(\tau-1)}]=0$ and $P_{\tau}\not\in \Stab^{(\tau-1)}$. Note that such decreases would correspond to loss of quantum information in the quantum error-correcting code interpretation. As we discussed in the previous subsection, this is not very likely to occur in the volume-law phase.

\subsection{Examples}

Below, we look at a few examples.

\begin{enumerate}[1)]
\item Assume that $P_{j}=Z_{j}$ for $j=1,\cdots, \tau$ ($\tau\leq n$). In this case, the stabilizer group is generated by
\begin{align}
\Stab^{(\tau)} = \langle Z_{1},\cdots, Z_{\tau} \rangle.
\end{align}
We also have 
\begin{align}
Z_{j} |\Psi(m)\rangle = m_{j} |\Psi(m)\rangle.
\end{align}
\item Assume that $n=3$. Also assume that $P_{1}=Z_{1}$, $P_{2}=X_{1}$, $P_{3}=Z_{2}$ and $P_{4}=X_{2}$. We then have
\begin{align}
\Stab^{(1)} = \langle Z_{1} \rangle, \quad \Stab^{(2)} = \langle X_{1} \rangle, \quad 
\Stab^{(3)} = \langle X_{1}, Z_{2} \rangle, \quad \Stab^{(4)} = \langle X_{1}, X_{2} \rangle
\end{align}
where anti-commuting generators are eliminated by adding $P_{2}$ and $P_{4}$. Eigenvalues depend only on $m_{2}$ and $m_{4}$:
\begin{align}
X_{1} |\Psi(m)\rangle = m_{2} |\Psi(m)\rangle \qquad X_{2} |\Psi(m)\rangle = m_{4} |\Psi(m)\rangle.
\end{align}

Logical operators can be found recursively 
\begin{equation}
\begin{split}
&\Logic^{(1)} = \langle Z_{1}, X_{2}, Z_{2}, X_{3}, Z_{3} \rangle, \qquad \Logic^{(2)} = \langle X_{1} ,X_{2}, Z_{2},  X_{3}, Z_{3} \rangle \\
&\Logic^{(3)} = \langle X_{1}, Z_{2}, X_{3}, Z_{3} \rangle, \qquad \Logic^{(4)} = \langle X_{1}, X_{2}, X_{3}, Z_{3} \rangle
\end{split}
\end{equation}
which are commutants of stabilizer groups. 

We can check that logical operators are null operators. For non-trivial logical operators $X_{3},Z_{3}$ in $\Logic^{(4)}$, we have 
\begin{align}
\C(X_{3}) =\C(Z_{3}) = (1,1,1,1). 
\end{align}
For stabilizer generators contained in $\Logic^{(4)}$, we have
\begin{align}
\C(X_{1}) = (-1, 1,1,1) = \E(P_{2}) \qquad \C(X_{2}) = (1,1,-1,1) = \E(P_{4}).
\end{align}
\item Assume that $n=2$. Also assume that $P_{1}=Z_{1}$, $P_{2}=Z_{2}$ and $P_{3}=X_{1}X_{2}$. We then have
\begin{align}
\Stab^{(1)} = \langle Z_{1} \rangle, \quad \Stab^{(2)} = \langle Z_{1}, Z_{2} \rangle, \quad 
\Stab^{(3)} = \langle Z_{1}Z_{2}, X_{1}X_{2} \rangle
\end{align}
where adding $P_{3}$ generated a larger stabilizer generator $Z_{1}Z_{2}$. We also have 
\begin{align}
Z_{1}Z_{2} |\Psi(m)\rangle = m_{1}m_{2} |\Psi(m)\rangle \qquad X_{1}X_{2} |\Psi(m)\rangle = m_{3} |\Psi(m)\rangle. 
\end{align}
We have
\begin{align}
\Logic^{(1)} = \langle Z_{1}, X_{2}, Z_{2} \rangle, \quad \Logic^{(2)} = \langle Z_{1}, Z_{2} \rangle, \quad 
\Logic^{(3)} = \langle Z_{1}Z_{2}, X_{1}X_{2} \rangle.
\end{align}
For operators in $\Logic^{(3)}$, we see
\begin{align}
\C(Z_{1}Z_{2}) = (1,1,1) \qquad \C(X_{1}X_{2}) = (-1,-1,1).
\end{align}
\end{enumerate}

\section{Distilling the Choi-Jamio\l{}kowski state}\label{sec:R-distillation}

In this section, we will present an algorithm to distill an entangled state from the system and the reference $R$ and show that it is identical to the Choi-Jamio\l{}kowski state of the underlying stabilizer code. 

\subsection{Reverse error vector}

The overall distillation procedure is graphically summarized as follows:
\begin{align}
 \figbox{1.5}{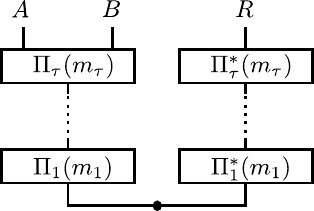}. \label{eq:MQC-summary-Ref}
\end{align}
where we perform projective measurements of complex conjugates $\Pi^{*}(m)$. We then apply some appropriate feedback operation on $R$. 

Recall that the original error vector $\E(P_{j})$ was constructed by looking at commutation relations with $P_{i}$ for $i<j$ in the past. Here, we instead need to introduce the reverse error vector $\E_{\rev}(P_{i})$ by looking at commutation relations with $P_{i}$ for $i>j$. In other words, we only look at commutations with respect to Pauli operators in the future~\footnote{
While we do not have an intuitive explanation for the need of reverse error vectors, one possible hint may be obtained by observing $\mathcal{C}(P_{j})=\mathcal{E}_{\rev}\cdot \mathcal{E}(P_{j})$.
A mathematical reason for considering the reverse error vector is presented in appendix~\ref{app:code}. 
}. Namely, we define
\begin{equation}
\begin{split}
&\E_{\rev}(P_{i})_{j} = 1 \qquad \qquad \qquad \qquad \qquad \qquad \ \ \ (j < i) \\
&\E_{\rev}(P_{i})_{j} = \pm 1 \qquad P_{i} P_{j} =  \pm P_{j} P_{i} \qquad \qquad (j \geq i).
\end{split}
\end{equation}
When $\E(P_{i})_{j}$ and $\E_{\rev}(P_{i})_{j}$ are interpreted as matrices, they are related by transpose, namely
\begin{align}
\E(P_{i})_{j} = \E_{\rev}(P_{j})_{i}.
\end{align}

Let us denote the group generated by reverse error vectors as 
\begin{align}
\E_{\rev} \equiv \big\langle \{ \E_{\rev}(P_{j})  \} \big\rangle.
\end{align}
In appendix~\ref{app:code}, we will prove the following lemma.

\begin{lemma}\label{lemma:reverse}
In the distillation algorithm from Eq.~\eqref{eq:MQC-summary-Ref}, measurement of $s (=m\cdot \overline{m})$ occurs if and only if 
\begin{align}
s \in \E_{\rev}.
\end{align}
\end{lemma}

This lemma suggests that, given $s=m\cdot \overline{m}$, one can always find a set of indices $\Lambda \subseteq \{1,\cdots,\tau\}$ such that
\begin{align}
s = \prod_{j \in \Lambda} \E_{\rev}(P_{j})
\end{align}
where $\prod$ represents component-wise multiplications of vectors.
The necessary feedback operation is given by
\begin{align}
P_{\Lambda} \equiv \prod_{j\in \Lambda} P_{j}.
\end{align}
We then have the following result.

\begin{lemma}\label{lemma:output-sys-ref-distillation}
The output of the aforementioned distillation algorithm for the system-reference entanglement is 
\begin{align}
\mathbb{E}\big(\sigma_{AB\overline{AB}}\big) = \frac{1}{N_{\mathcal{S}}} \sum_{P \in \mathcal{S} } | P \rangle \langle P |
\end{align}
where $N_{\mathcal{S}}$ is the number of elements in $\mathcal{S}$. 
\end{lemma}

The proof of this lemma is presented in appendix~\ref{app:code}.

\subsection{Choi-Jamio\l{}kowski state}

Let $P \in \mathcal{L}$ be a Pauli operator in the logical operator group. Then the output state from the distillation algorithm satisfies 
\begin{align}
\Tr\Big[ (P \otimes P^*)\mathbb{E}( \sigma_{AB\overline{AB}}) \Big] = 1
\end{align}
since stabilizer generators commute with $P$. This implies that the output state $\mathbb{E}( \sigma_{AB\overline{AB}})$ satisfies
\begin{align}
\big\langle \overline{Z_{j}} \otimes \overline{Z^*_{j}} \big\rangle = 1 \qquad 
\big\langle \overline{X_{j}} \otimes \overline{X^*_{j}} \big\rangle = 1 \qquad  j=1,\cdots, k. \label{eq:logical_correlation}
\end{align}
Hence, $k$ EPR pairs can be distilled from $AB$ and $\overline{AB}$, and $\overline{X_{j}},\overline{Z_{j}}$ for $j=1,\cdots, k$ transform encoded quantum information by acting as Pauli $X$ and $Z$ operators on logical qubits. On the other hand, we have
\begin{align}
\big\langle \overline{Z_{j}} \otimes \overline{Z^*_{j}}\big\rangle = 1 \qquad  \big\langle \overline{X_{j}} \otimes \overline{X^*_{j}} \big\rangle = 0 \qquad j=k+1,\cdots, n. 
\end{align}
Here $\overline{X_{j}}$ for $j=k+1,\cdots, n$ are defined as anti-commuting partners of $\overline{Z_{j}}$. This suggests that $AB$ and $\overline{AB}$ retains classical correlation with respect to eigenvalues of $\overline{Z_{j}}$ and $\overline{Z^*_{j}}$.

The classical correlation in $\mathbb{E}( \sigma_{AB\overline{AB}})$ results from averaging over $m$. If one looks at the distilled state for each $m$, we find that 
\begin{align}
\langle \overline{Z_{j}} \otimes I \rangle = \langle I \otimes \overline{Z^*_{j}} \rangle  = 1 \qquad \mbox{for}\quad \sigma_{AB\overline{AB}}(m).
\end{align}
Here we assumed that $\overline{Z_{j}}$ is properly relabelled so that the output state is stabilized by $\mathcal{S}(m)$ where the expectation values are taken with respect to $\sigma_{AB\overline{AB}}(m)$. Note that Eq.~\eqref{eq:logical_correlation} holds for each $\sigma_{AB\overline{AB}}(m)$ as well. Hence, we can conclude that $\sigma_{AB\overline{AB}}(m)$ is nothing but the Choi-Jamio\l{}kowski state of a stabilizer code with the stabilizer group $\mathcal{S}(m)$.

Applying this version of the distillation algorithm to the Gullans-Huse proposal will distill an EPR pair in the encoded basis states instead of a pair of qubits. 

\section{State-independent entanglement structure}
\label{sec:state-independence}

We have developed a theoretical framework to study the entanglement structure of monitored quantum circuits and derived several rigorous results. In the remainder of the paper, we discuss its implications on the physics of monitored quantum circuits in the volume-law phase.

In this section, we argue that the entanglement structure of a monitored quantum circuit changes drastically when the subsystem size $A$ exceeds a certain critical size, which can be identified as the code distance $d_{\text{code}}$. While we do not focus on specific models of monitored circuits, it will be useful to imagine random monitored Clifford circuits in the volume-law phase which has been running for longer than the entanglement equilibrium time to develop the volume-law entanglement, but shorter than the exponentially long quantum memory time.

\subsection{Decoupling and state-independence}

So far we have used a maximally mixed state as an initial state of monitored quantum circuits (or equivalently, we have appended the entangled reference system $R$). A naturally arising question concerns the entanglement structure when a pure state, instead of a maximally mixed state, is prepared as an initial state. Here we argue that entanglement between two subsystems $A$ and $B$ is largely independent of the choice of initial states as long as the size of the smaller subsystem $A$ is below a certain critical size which plays the role of the code distance $d_{\text{code}}$ of a monitored quantum circuit.

Let us begin by recalling the notion of decoupling (see~\cite{Hayden:2008aa} for rigorous arguments). Assume that a smaller subsystem $A$ is strongly entangled with $B$, namely
\begin{align}
I_{(A,B)} \approx 2S_{A}.
\end{align}
Here note that $I_{(A,B)}\leq 2S_{A}$. Recalling $I_{(A,B)} + I_{(A,R)} = 2S_{A}$ for a pure state on $ABR$, we then notice that the subsystem $A$ is almost completely decoupled from the reference $R$ with $I_{(A,R)}\approx 0$~\footnote{A conventional definition of decoupling is $\rho_{AR}\approx \rho_{A} \otimes \rho_{R}$. Here we use the word ``decoupling'' in a loose sense by referring to the situation with small $I_{(A,R)}$. To obtain a useful quantitative decoupling inequality, it is often more convenient to use R\'{e}nyi generalizations of mutual information which can be accessed from OTOCs~\cite{Yoshida:2019aa}.}. This suggests that entanglement between $A$ and $B$ are largely independent of the initial state of the circuit. To be concrete, let us pick some pure state, such as product states or Haar random states, as an initial state of a monitored quantum circuit. This situation can be realized by performing a projective measurement on the reference system $R$. Namely, if we project $R$ onto $|\psi^{*}\rangle$, then the initial state on the system $AB$ will be set to $|\psi\rangle$. Since the reference $R$ is decoupled from $A$, quantum operations on $R$ cannot make a significant influence on the entanglement between $A$ and $B$. As such, the entanglement between $A$ and $B$ is largely independent of initial states~\footnote{Since $I_{(A,R)}$ is not exactly zero, there may exist fine-tuned initial states which generate atypical entanglement.}. 

Now, recall that a monitored quantum circuit in the volume-law phase can be interpreted as a quantum error-correcting code that is robust against local projective measurements. This suggests that logical operators of the code cannot be supported on a small subsystem. Namely, if the subsystem $A$ is smaller than the code distance $d_{\text{code}}$, we expect to have 
\begin{align}
I_{(A,R)} \approx 0 \qquad \text{when} \quad |A| < d_{\text{code}}
\end{align}
since, otherwise, an approximate logical operator can be constructed on $A$. The above equation can be interpreted as a definition of the approximate code distance. Indeed, it is useful to recall that conditions for approximate error-correction can be expressed in terms of the coherent information, which can be also interpreted as the conditional entropy $S_{A|R}$~\cite{Schumacher:2002aa}. This observation suggests that the entanglement structure of a monitored quantum circuit is largely independent of initial states up to the size scale of the code distance $d_{\text{code}}$ due to decoupling.

\subsection{Estimate of the code distance}

The code distance $d_{\text{code}}$ depends on the specifics of the model of interest. Here we argue that, for a monitored quantum circuit in the volume-law phase with an exponentially long memory time, the code distance $d_{\text{code}}$ scales polynomially with respect to the system size $n$~\footnote{From the conventional wisdom on phase transitions, it will be natural to expect an exponential memory time when the system is away from the criticality of the entanglement phase transition.}. 

We begin with the cases of Clifford circuits. In a stabilizer quantum error-correcting code, there always exists a subsystem $A$ consisting of $d_{\text{code}}$ qubits which support a non-trivial Pauli logical operator $\ell = P_{A_{1}} \otimes P_{A_{2}} \otimes \cdots \otimes P_{A_{d_{\text{code}}}}$. Suppose that one performs projective measurements with some finite probability $p>0$. The quantum information associated with the logical operator $\ell$ will be lost when one accidentally measures $P_{A_{1}} \otimes P_{A_{2}} \otimes \cdots \otimes P_{A_{d_{\text{code}}}}$. This event occurs with probability 
\begin{align}
\left(\frac{p}{3}\right)^{d_{\text{code}}}. \label{eq:probability-estimate}
\end{align}
Hence a monitored circuit will lose a piece of quantum information in one unit of time at least with the probability in Eq.~\eqref{eq:probability-estimate}. This suggests that the quantum memory time is upper bounded by 
\begin{align}
t_{\text{memory}} \lessapprox \left(\frac{3}{p}\right)^{d_{\text{code}}}
\end{align}
which sets a lower bound on $d_{\text{code}}$:
\begin{align}
\log ( t_{\text{memory}} ) \lessapprox d_{\text{code}} \log\Big(\frac{3}{p}\Big).
\end{align}
Therefore, in order to have an exponential quantum memory time, the code distance $d_{\text{code}}$ needs to scale polynomially with respect to the total system size $n$~\footnote{Here, by an exponential memory time, we mean that $t_{\text{memory}}$ grows as $\simeq \exp(n^\gamma)$ with $\gamma>0$.}:
\begin{align}
d_{\text{code}} = \text{Poly}(n).
\end{align}
It is worth noting that the above analysis only gives a lower bound on $d_{\text{code}}$, and hence does not give a sufficient condition for an exponential memory time.

For non-Clifford circuits, logical operators cannot be written as a tensor product of Pauli operators, so the above argument is not readily applicable. Here, we argue that a similar lower bound still applies based on the relation between random Clifford and Haar random unitary operators. Recall that, in monitored Clifford circuits, the loss of quantum information occurs in a discrete manner in time steps. Namely, when the mutual information $I(AB,R)$ decreases, its value drops by an integer, while for other times, $I(AB, R)$ may stay constant. Only when one considers the behavior of $I(AB, R)$ averaged over all the statistical realizations of random Clifford circuits, continuous decays by non-integer values can be found. This is because, for Clifford circuits, the sample-to-sample variance is large. For monitored quantum circuits driven by non-Clifford dynamics, we expect that the decrease of $I(AB, R)$ will be continuous. This intuition is based on an observation that the statistical average for random Clifford dynamics converges to the result from a single realization of Haar random circuits due to the concentration of measure (which can be verified by computing the variance). Hence, we conclude that, for non-Clifford circuits, the mutual information $I(AB,R)$ will decrease continuously over the period of $t_{\text{memory}}$, and as such, the code distance $d_{\text{code}}$ should be polynomial in $n$.

Finally, let us note that Li and Fisher investigated the code distance~\footnote{Strictly speaking, Li and Fisher studied the contiguous code distance, instead of the conventional code distance, by looking at single intervals only.} of one-dimensional random monitored Clifford circuits and numerically obtained an estimate of $d_{\text{code}} \sim n^{0.36}$~\cite{Li:2021aa} (see~\cite{Gullans:2020aa} for an estimate from an earlier work which qualitatively match with this estimate). 

With these observations and previous results, we conclude that the entanglement structure of a monitored quantum circuit in the volume-law phase is independent of initial states up to a $\text{Poly}(n)$ size scale which is of the order of the code distance. 

\subsection{Measurement history dependence}

Finally, we discuss how the entanglement depends on measurement outcomes in the past. Namely, we will argue that measurements in the distant past are largely irrelevant to the entanglement between two subsystems $A$ and $B$ when $A$ is smaller than $d_{\text{code}}$.

Let us split the measurement operator $\Pi(m)$ into two parts $\Pi(m)=\Pi_{\text{recent}} (m_{\text{recent}}) \Pi_{\text{past}} (m_{\text{past}})$ where
\begin{equation}
\begin{split}
\Pi_{\text{recent}} (m_{\text{recent}}) &\equiv \Pi_{\tau}(m_{\tau}) \cdots \Pi_{\tau - \Delta \tau}(m_{\tau - \Delta \tau})\\
\Pi_{\text{past}} (m_{\text{past}}) &\equiv \Pi_{\tau - \Delta \tau -1}(m_{\tau - \Delta \tau - 1}) \cdots  \Pi_{1}(m_{1}).
\end{split}
\end{equation}
The output wavefunction can be expressed in the following manner
\begin{align}
\figbox{1.5}{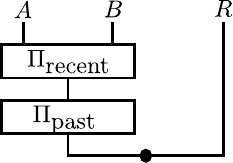} \ =\ \figbox{1.5}{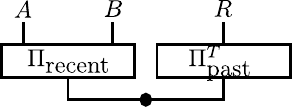} 
\end{align}
where we moved $\Pi_{\text{past}} (m_{\text{past}})$ to the right hand side by taking transpose. Written in this form, we can interpret the above wavefunction as an output wavefunction from $\Pi_{\text{recent}} (m_{\text{recent}})$ whereas $\Pi_{\text{past}} (m_{\text{past}})$ acts as a projection on the reference $R$. (In other words, the portion of $\Pi_{\text{past}} (m_{\text{past}})$ can be interpreted as the initial state of $\Pi_{\text{recent}} (m_{\text{recent}})$).

Here we assume that $\Delta \tau$ is longer than the entanglement equillibrium time so that the subsystem $A$, which is smaller than the code distance $d_{\text{code}}$, is decoupled from the reference $R$. From arguments in previous subsections, we find that the entanglement between $A$ and $B$ is immune to projective measurements on the reference $R$. Hence, we can conclude that the measurements in the distant past $\Pi_{\text{past}} (m_{\text{past}})$ does not affect the entanglement between $A$ and $B$. This suggests that the entanglement between $A$ and $B$ can be distilled without knowing $m_{\text{past}}$. As such, the entanglement structure below the $d_{\text{code}}$ scale depends only on measurements that occurred within the entanglement equilibrium time~\footnote{For one-dimensional random monitored circuits with product initial states (\emph{e.g.} $|0\rangle^{\otimes n}$), it will take $O(n)$ time in order to reach the entanglement equilibrium with the volume-law entanglement~\cite{Skinner:2019aa}. A recent study in~\cite{Li2021}, however, seems to suggest that it will take only $O(n^{\frac{2}{3}})$ time in order to reach the steady value of the entanglement entropy if one starts with a maximally mixed state instead of product states. This is due to the observation that the entangling minimal surface of a subsystem $A$ extends into the bulk with the depth $\sim |A|^{\frac{2}{3}}$ only, instead of $\sim|A|$. As such, the decoupling with $I(A,R)\approx 0$ will occur in $O(n^{\frac{2}{3}})$ time instead of $O(n)$. We speculate that this is due to a possibility that the size of stabilizer generators may grow faster than linear in the presence of projective measurements where multiple stabilizer generators may need to be combined to form new stabilizer generators. This will not lead to any causality violation since the verification of entanglement needs to know the measurement outcomes which can travel only at the speed of light.}.

\section{State-dependent entanglement structure}\label{sec:state-dependence}

Once the subsystem $A$ becomes larger than the code distance $d_{\text{code}}$, the mutual information $I_{(A,R)}$ may take a non-zero value. In this case, the entanglement structure between $A$ and $B$ will be dependent on the initial states of the monitored quantum circuit as well as measurement outcomes in the distant past. Here, we present a heuristic argument concerning how the mutual information $I_{(A,B)}$ changes by preparing a generic pure state as an initial state instead of the maximally mixed state.

\subsection{Entanglement swapping by random projection}

In order to gain some insight, it is useful to consider a simplified toy model of the entanglement structure involving $A,B,R$ as shown below:
\begin{align}
\figbox{1.5}{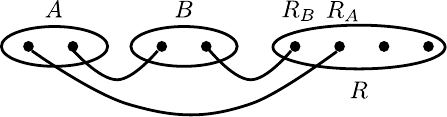}\ .
\end{align}
where bipartite entanglement (\emph{e.g.} EPR pairs) are distributed among $A,B,R$. The bipartite entanglement between $A$ and $B$ represents the state-independent entanglement which exists below the $d_{\text{code}}$ scale whereas the $A-R$ and $B-R$ entanglement are associated with the encoding of logical qubits, and can be accessed only above the $d_{\text{code}}$ scale. Here $R_{A}$ and $R_{B}$ represent degrees of freedom which are entangled with $A$ and $B$ respectively.

Let us think of projecting $R$ onto some pure state $|\psi\rangle_{R}$. If $|\psi\rangle$ is a product state on $R_{A}\otimes R_{B}$, the projection will not generate any additional entanglement, and the value of $I_{(A,B)}$ remains unchanged. On the other hand, if $|\psi\rangle$ is entangled across $R_{A}$ and $R_{B}$ (\emph{e.g.} an EPR pair on $R_{A}\otimes R_{B}$), the projection will lead to additional entanglement:
\begin{align}
\figbox{1.5}{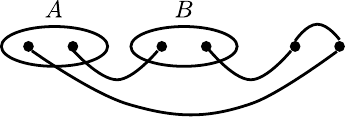}\ . \label{eq:D-Brane}
\end{align}
Namely, the original entanglement between $A$ and $R$ is merged with the entanglement between $B$ and $R$, and then contributes as additional entanglement between $A$ and $B$. Note that this additional entanglement depends on how $A,B$ were entangled with $R$, as well as the choice of the entangled state $|\psi\rangle_{R}$ on $R$. 

This mechanism can be interpreted as the quantum teleportation (or the entanglement swapping). Namely, Bell measurements on $R_{A}$ and $R_{B}$ can send $R_{A}$ to a subsystem $B$ by using the $B-R_{B}$ entanglement as a resource. This forces the qubits on $A$, which were initially entangled with $R_{A}$, to be entangled  with $B$. In other words, the $A-R$ entanglement was swapped to become the $A-B$ entanglement.

As this observation suggests, collapsing $R$ into an entangled state tends to increase the mutual information $I_{(A,B)}$. One can make this observation more rigorous by considering a projection onto a Haar random state on $R$. Namely, one can show that the R\'{e}nyi-$2$ entanglement entropy $S_{A}^{(2)}$ does not change much after projecting $R$ onto a random state (assuming $A$ is the smaller subsystem). Consider the following output state with Haar random initial state $|\psi\rangle$:
\begin{align}
\frac{1}{\sqrt{\Prob(m)}} \ \figbox{1.5}{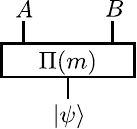}
\end{align}
where the numerical factor $\frac{1}{\sqrt{\Prob(m)}}$ achieves approximately proper normalization. Let us denote the density matrix of the above wavefunction by $\rho_{|\psi\rangle\langle \psi|}$. Then we have 
\begin{align}
\Tr({\rho_{A}}_{|\psi\rangle\langle \psi|}^2)  = \frac{1}{\Prob(m)^2} \ \figbox{1.5}{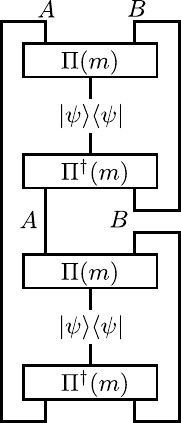}
\end{align}
Taking the Haar average leads to 
\begin{align}
\int d|\psi\rangle \Tr({\rho_{A}}_{|\psi\rangle\langle \psi|}^2) = \frac{d}{d+1}\Big( \Tr(\rho_{A}^2) + \Tr(\rho_{B}^2)  \Big)
\end{align}
where $\rho_{A}$ and $\rho_{B}$ are defined for the original output wavefunction that includes the entangled reference $R$. 

Here we assumed that $A$ is the smaller subsystem. Hence, it is natural to assume  
\begin{align}
\Tr(\rho_{B}^2)\ll \Tr(\rho_{A}^2).
\end{align}
Recalling that $d=2^n$, we find that $S_{A}^{(2)}$ stays approximately the same after projecting $R$ onto Haar random states:
\begin{align}
{S_{A}}_{|\psi\rangle\langle \psi|}^{(2)} \approx S_{A}^{(2)}.
\end{align}
While this analysis computed R\'{e}nyi-$2$ entropy, we expect that the entanglement entropy behaves similarly. This suggests that the value of $I_{(A,B)}$ will increase roughly by $I_{(A,R)}$, namely
\begin{align}
\text{$I_{(A,B)}$ with random $R$ projection} \approx \text{$I_{(A,BR)}$ with no $R$ projection}.
\end{align}
Hence, after the random projection on $R$, the subsystem $A$ will be entangled with $B$ without losing its initial entanglement with $BR$. This increase of $I_{(A,B)}$ can be viewed as the entanglement swapping by a random projection. 

\begin{figure}
\centering
\includegraphics[width=0.45\textwidth]{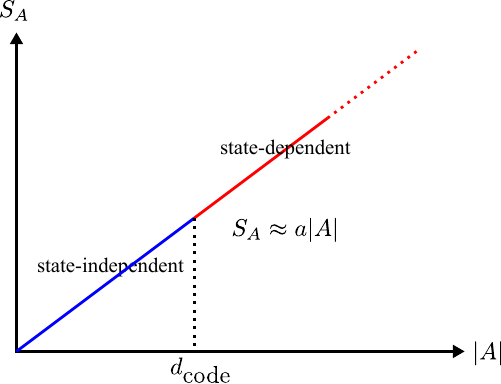}
\caption{
The volume-law scaling in the volume-law phase. The overall behavior of $S_{A}$ does not depend on the initial states (except atypical ones) due to the entanglement swapping. Across the $d_{\text{code}}$ scale, however, the entanglement changes from being state-independent to state-dependent. We expect that the complexity of the entanglement verification changes drastically. 
}
\label{fig-plot}
\end{figure}

We speculate that the above conclusion for Haar random initial states also applies to the cases when product states are chosen as initial states, since the degrees of freedom $R_{A}$ and $R_{B}$ will be non-local on the reference system $R$, and thus projecting $R$ onto a product state has an effect of projecting $R_{A}$ and $R_{B}$ (as well as their complementary systems) onto entangled states. Hence we expect that the value of the entanglement entropy $S_{A}$ is largely independent of the initial states (except fine-tuned ones), as in Fig.~\ref{fig-plot}. This observation is consistent with previous numerical and analytical results, see~\cite{Li:2021aa} for instance. This provides an important caution that the volume-law scaling of the entanglement entropy $S_{A}\approx a |A|$ is too crude to this subtle, yet important difference of the entanglement structure below and above the code distance scale. In the next section, we will argue that the subleading contribution to the entanglement entropy probes coding properties of a monitored quantum circuit.

\subsection{On complexity of entanglement verification}

One salient feature of the state-independent entanglement below the $d_{\text{code}}$ scale is that its verification does not require knowledge of measurement outcomes in the distant past. This suggests that the quantum complexity of the entanglement verification may change drastically across the $d_{\text{code}}$ scale. Indeed, for Clifford monitored circuits, when a subsystem $A$ is smaller than $d_{\text{code}}$, the distillation algorithm can run in a time scale comparable to the entanglement equilibrium time which is polynomial in the system size. For subsystems larger than $d_{\text{code}}$, however, the algorithm needs to know measurement outcomes in the distance past as well as the initial state. This suggests that the distillation complexity can be large if the circuit has been running for much longer than the entanglement equilibrium time~\footnote{While an arbitrary Clifford operator can be implemented efficiently on a quantum computer, we expect that processing exponentially many measurement outcomes cannot be done efficiently.}. As such, for monitored Clifford circuits, there will be a ``phase transition'' of the entanglement verification complexity across the $d_{\text{code}}$ scale~\cite{Yoshida:2020aa}~\footnote{
If one hopes that the entanglement in monitored quantum systems would ever be relevant to physically observable phenomena, it must be verifiable. In this regard, one may speculate that the state-independent entanglement below the $d_{\text{code}}$ scale will be responsible for such a phenomena (if exists).}.

\subsection{Does measurement destroy entanglement?}

Discussions so far reveal a certain tension between the conventional understanding of the physics of monitored quantum circuits and the role of projective measurements concerning the emergence of the volume-law entanglement. It is commonly believed that projective measurements in monitored quantum circuits lead to decoherence which \emph{destroys} entanglement. Namely, the conventional understanding of the emergence of the volume-law entanglement is that the effect of scrambling dynamics, which create entanglement, can outperform the decoherence from local projective measurements. This intuition can be made concrete by recalling the simplified toy model of monitored quantum circuits with intermittent projective measurements due to Choi \textit{et al.}~\cite{Choi:2020aa}. In this toy model, the system is separated into groups of multiple qubits where neighboring groups of qubits interact with each other via random unitaries. Once neighboring groups of qubits are throughly mixed, local projective measurements are performed. In this toy model, random unitary dynamics can encode preexisting entanglement into subspaces of quantum error-correcting codes which protect the volume-law entanglement from local projective measurements.

As the above observation from the toy model suggests, projective measurements appear to destroy entanglement. However, this lesson should be understood with caution. As we have discussed throughout this paper, two subsystems $A$ and $B$ can be entangled in a state-independent manner due to the decoupling phenomena induced by projective measurements. Specifically, let us consider the case where the initial state is a maximally mixed state $\mu_{A}\otimes \mu_{B} = \frac{1}{d} I_{A}\otimes I_{B}$. Observe that, if no measurements were performed, then the system would remain unentangled because $\mu_{A}\otimes \mu_{B}$ is invariant under the action of any unitary operator. Once projective measurements are performed, however, the output quantum state can start to develop entanglement between $A$ and $B$. Hence, in this case with a maximally mixed initial state, local projective measurements \emph{create} entanglement, instead of destroying it. One might think that this has to do with the special case of a maximally mixed initial state, but the entanglement structure between $A$ and $B$ is independent of the initial states. 

A naturally arising question then is whether projective measurements destroy or create entanglement. The resolution of this apparent tension is immediate from discussions in previous subsections. Below the code distance scale, the entanglement structure is independent of the initial states, and thus projective measurements are indeed creating entanglement via the decoupling phenomena. Above the code distance scale, on the other hand, the subsystem in the output wavefunction starts to be correlated with the initial states. In this regime, it is reasonable to view the monitored circuit as an encoding into a quantum error-correcting code which protects entanglement from projective measurements which would destroy entanglement. Hence, projective measurements can create or destroy entanglement, depending on the size scale of interest. Furthermore, from this perspective, we argue that the toy model from~\cite{Choi:2020aa} captures the coarse-grained physics of monitored quantum circuits above the code distance scale.

\section{Code distance from sub-leading entropy}\label{sec:gamma}

Observations from the previous sections resolve a certain puzzle concerning the sub-leading contribution to the volume-law entanglement entropy in a monitored quantum circuit. 

Several previous works have conjectured that, in the volume-law phase of a monitored quantum circuit, there will be a logarithmic sub-leading contribution to the volume-law entanglement~\cite{Fan:2021aa, Li:2019aa}. Namely, for one-dimensional circuits, the following form of asymptotic entanglement scaling has been conjectured:
\begin{align}
S_{A} = a L_{A} + c \log L_{A}.
\end{align}
Certain physical arguments to explain the origin of the logarithmic term have been presented in~\cite{Fan:2021aa, Li:2019aa} based on size distributions of stabilizer generators and an entropy drop via projective measurements. However, Li and Fisher, who numerically studied a one-dimensional Clifford circuit in a later work~\cite{Li:2021aa}, have found that there is another sub-leading contribution, namely
\begin{align}
S_{A} = a L_{A} + b {L_{A}}^{\gamma} + c \log L_{A} \label{eq:subleading}
\end{align}
with some exponent $\gamma\approx 0.38$. Here $L_{A}$ is the length of $A$. 

\begin{figure}
\centering
\includegraphics[width=0.55\textwidth]{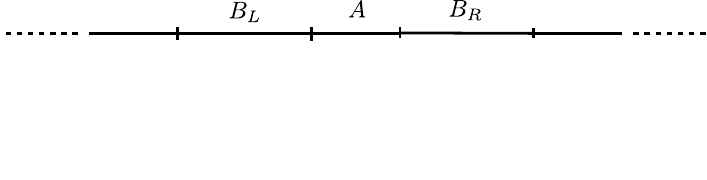}
\caption{
An heuristic argument showing $\gamma = \gamma_{\text{code}}$.
}
\label{fig-subleading}
\end{figure}

Here we present a heuristic argument showing that the sub-leading term $\sim{L_{A}}^{\gamma}$ results from coding properties of the underlying monitored quantum circuit~\footnote{In a recent work~\cite{Li2021}, Li, Vijay and Fisher utilized an effective theory description of one-dimensional monitored quantum circuits and attributed the origin of the $\sim{L_{A}}^{\gamma}$ term as a fluctuation of the entangling surface. }. Namely, we claim that the exponent $\gamma$ is equal to the exponent for the code distance $d_{\text{code}} \sim L^{\gamma_{\text{code}}}$:
\begin{align}
\gamma = \gamma_{\text{code}}. 
\end{align}
Let us pick three neighboring subsystems $B_{L}$, $A$ and $B_{R}$ such that $B=B_{L}\cup B_{R}$ surrounds $A$  (Fig.~\ref{fig-subleading}). Let $L_{A}$ and $L_{B}$ be the lengths of $A$ and $B_{L}, B_{R}$ respectively. Let us compute the mutual information $I_{(A,B)}$ by using the asymptotic entanglement scaling formula in Eq.~\eqref{eq:subleading}. We have 
\begin{align}
S_{A} \approx a L_{A} + b {L_{A}}^{\gamma} \qquad S_{B_{L}} \approx S_{B_{R}} \approx a L_{B} + b {L_{B}}^{\gamma}.
\end{align}
We also have
\begin{align}
S_{AB} \approx a (L_{A}+2L_{B}) +  b (L_{A}+2L_{B})^{\gamma}.
\end{align}
Finally, we need to compute $S_{B}$. At this moment, let us assume that $B_{L}$ and $B_{R}$ are not entangled with each other since $B_{L}$ and $B_{R}$ are separated by $A$. (We will return to this assumption in a few paragraphs). Then, we have
\begin{align}
S_{B} \approx S_{B_{L}} + S_{B_{R}} \approx 2a L_{B} + 2b {L_{B}}^{\gamma}. \label{eq:assumption-correlation}
\end{align}
Using these asymptotic estimates, we obtain 
\begin{align}
I_{(A,B)} \approx b L_{A}^{\gamma} + 2b L_{B}^{\gamma} -  b (L_{A}+2L_{B})^{\gamma}
\end{align}
where the volume terms cancel with each other. 

Let us fix $L_{A}$ and increase $L_{B}$. As $L_{B}$ becomes larger than $L_{A}$, the above estimate can be further approximated by 
\begin{align}
I_{(A,B)} \approx b L_{A}^{\gamma} + 2b L_{B}^{\gamma} - b (2L_{B})^{\gamma} \left( 1 + \gamma \frac{L_{A}}{2L_{B}} \right) \approx  b( 2 - 2^{\gamma} )  L_{B}^{\gamma}.  \label{eq:mutual}
\end{align}
So, $I_{(A,B)}$ grows with the exponent $\gamma$ as we increase $L_{B}$. However, $I_{(A,B)}$ is upper bounded by $2S_{A}\approx 2a L_{A}$, so we expect that $I_{(A,B)}$ will get saturated when 
\begin{align}
L_{A} \approx L_{B}^{\gamma}
\end{align}
where we ignored the constants $a,b$. Hence, as long as $L_{B} \gg L_{A}^{\frac{1}{\gamma}}$, two subsystems $A$ and $B$ are nearly maximally entangled, and $A$ is decoupled from the reference $R$. In summary, we have obtained the following estimate:
\begin{equation}
\begin{split}
I_{(A,B)}  &\approx L_{B}^{\gamma} \qquad (L_{B} \lessapprox L_{A}^{\frac{1}{\gamma}}) \\
&\approx L_{A} \qquad (L_{B} \gtrapprox L_{A}^{\frac{1}{\gamma}}).
\end{split}
\end{equation}

Now we think of increasing both $L_{A}$ and $L_{B}$. Recall that the value of $L_{B}$ is upper bounded by the system size $L$. Then, if $L_{A} \gtrapprox L^{\gamma}$, one cannot take a large enough subsystem $B$ such that $I_{(A,R)}=0$. Namely, we expect that $A$ will be entangled with the reference $R$ once $L_{A}$ becomes larger than $L^{\gamma}$. Hence, we can conclude that the code distance scales as
\begin{align}
d_{\text{code}} \approx L^{\gamma} \qquad \mbox{and} \qquad \gamma = \gamma_{\text{code}}.
\end{align}
For one-dimensional random monitored Clifford circuits, Li and Fisher numerically estimated $\gamma \approx 0.36$ and $\gamma_{\text{code}} \approx 0.38$ which is consistent with this argument.

Since $I_{(A,B)}$ is upper bounded by $2S_{A}$, our estimate of the mutual information in Eq.~\eqref{eq:mutual} is valid only for $L_{B} \lessapprox {L_{A}}^{\frac{1}{\gamma}}$. When $L_{B} \gtrapprox{L_{A}}^{\frac{1}{\gamma}}$, we expect that our assumption of $S_{B} \approx S_{B_{L}} + S_{B_{R}}$ in Eq.~\eqref{eq:assumption-correlation} becomes invalid. In order to verify this expectation, we will evaluate the mutual information between $B_{L}$ and $B_{R}$. Let us begin by computing the mutual information between $B_{L}$ and $AB_{R}$. We have
\begin{align}
I_{(B_{L},AB_{R})} = S_{B_{L}} + S_{AB_{R}} - S_{B_{L}AB_{R}} \approx bL_{B}^{\gamma} + b(L_{A} + L_{B})^{\gamma} - b(L_{A} + 2L_{B})^{\gamma} \approx b(2-2^{\gamma})L_{B}^{\gamma}. \label{eq:mutual-BAB}
\end{align}
The mutual information $I_{(B_{L},B_{R})}$ can be lower bounded by using the generic upper bound on the conditional mutual information:
\begin{align}
2 S_{A} \geq I_{(B_{L}, A B_{R})} - I_{(B_{L}, B_{R})}.
\end{align}
This leads to 
\begin{align}
I_{(B_{L},B_{R})} \gtrapprox b(2-2^{\gamma})L_{B}^{\gamma} - 2aL_{A}. \label{eq:lower-bound-mutual}
\end{align}

The lower bound Eq.~\eqref{eq:lower-bound-mutual} becomes non-trivial for $L_{B} \gtrapprox L_{A}^{\frac{1}{\gamma}}$, which is exactly when we expect that the assumption of $S_{B} \approx S_{B_{L}} + S_{B_{R}}$ starts to become invalid due to the saturation of $I_{(A,B)}$. Here we expect that this lower bound is saturated
~\footnote{For Clifford circuits, the saturation of Eq.~\eqref{eq:lower-bound-mutual} can be argued by studying the sizes of stabilizer generators. In order to compute $I(B_{L},B_{R})$, we need to find the number of stabilizer generators which are supported non-locally over  $B_{L}$ and $B_{R}$. Eq.~\eqref{eq:mutual-BAB} suggests that there are $b(2-2^{\gamma})L_{B}^{\gamma}$ independent stabilizer generators which are supported non-locally over $B_{L}$ and $AB_{R}$. Given such a non-local stabilizer over $B_{L}$ and $AB_{R}$, we look at the profile of Pauli operators on $A$. If its support on $A$ belongs to the local stabilizer group $\mathcal{S}_{A}$ on $A$, the stabilizer generator can be brought into a form non-locally supported over $B_{L}$ and $B_{R}$, and hence it will make a contribution to $I(B_{L},B_{R})$. By noting that the profile of operators on $A$ must commute with $\mathcal{S}_{A}$, one notices that the number of such stabilizer generators can be lower bounded by $b(2-2^{\gamma})L_{B}^{\gamma} - 2S_{A}$, which is identical to Eq.~\eqref{eq:lower-bound-mutual}. The asymptotic saturation of this inequality can be argued by assuming that the Pauli operator profile of these stabilizer generators are random (with the constraint that they commute with $\mathcal{S}_{A}$).}
. Hence we obtain
\begin{equation}
\begin{split}
I_{(B_{L},B_{R})}  &\approx 0 \ \ \qquad (L_{B} \lessapprox L_{A}^{\frac{1}{\gamma}}) \\
&\approx L_{B}^{\gamma} \qquad (L_{B} \gtrapprox L_{A}^{\frac{1}{\gamma}}). 
\end{split}
\end{equation}
Here, strictly speaking, $I_{(B_{L},B_{R})}  \approx 0 $ means that $I_{(B_{L},B_{R})}$ is smaller than $L_{B}^{\gamma}$ in an asymptotic sense.  

\section{Relation to black hole physics}\label{sec:BH}

In this section, we establish a connection between monitored quantum circuits and black hole physics. Let us begin by arguing that monitored quantum circuits can be viewed as the Hayden-Preskill recovery problem running backward in time~\cite{Beni18}. 

The Hayden-Preskill recovery problem asks whether a piece of quantum information thrown into an old black hole, which is maximally entangled with the early radiation, can be retrieved by having access to both the early and late radiations. Information theoretically, this problem can be formulated as the following wavefunction
\begin{align}
 \figbox{1.5}{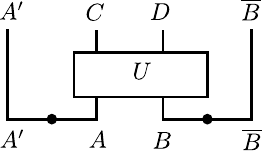}. \
\end{align}
where the old black hole is modelled as $n_{B}$ copies of EPR pairs on $B$ and $\overline{B}$ with $\overline{B}$ being the early radiation. The infalling quantum state is represented by EPR pairs on $A$ and $A'$ where $A'$ plays the role of the reference system. The system evolves by some unitary operator $U$, and $C$ and $D$ represent the remaining black hole and the late radiation respectively. In an information theoretic language, the Hayden-Preskill recovery problem asks whether quantum entanglement can be distilled from $A'$ and $\overline{B}D$.

Hayden and Preskill pointed out that the information is recoverable as long as $n_{D}\gtrapprox n_{A}$ when the dynamics $U$ is a Haar random unitary operator~\cite{Hayden07}. Later, it has been found that the information is recoverable when the black hole's dynamics is scrambling as quantified by OTOC functions~\cite{Yoshida:2017aa, Hosur:2015ylk, Roberts:2017aa}. Since the black hole scrambles quantum information, this result provides a formal proof that information can indeed leak out from an old black hole due to scrambling dynamics. Several concrete methods of retrieving quantum information from an old black hole have been proposed~\cite{Yoshida:2017aa, Gao:2017aa, Traversable2017, Brown:aa, Nezami:aa, Gao:aa, Schuster:aa}.

Here, instead of collecting the late Hawking radiations, let us think of performing projective measurements on late radiations in a continuous manner. This can be schematically represented as follows:
\begin{align}
 \figbox{1.5}{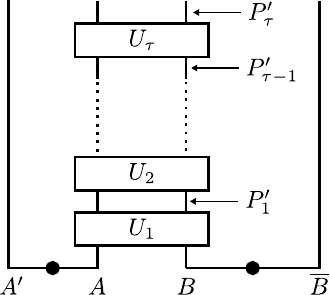} \ .
\end{align}
The central question is whether the quantum information is recoverable from the early radiation $\overline{B}$ as a result of projective measurements or not. In other words, we are interested in whether $A$ and $B$ are entangled or not. 

One may be able to see the similarity between this quantum circuit and the monitored circuit. Let us think of ``turning'' the diagram upside down so that the time flows backward (from the up to the bottom) and two subsystems $AB$ become the output of the quantum circuit. Then, one can see that the Hayden-Preskill recovery problem with continuous measurement is identical to the entanglement distillation problem in the monitored quantum circuit. Here the subsystem $A$ and $B$ in monitored quantum circuits correspond to the infalling quantum state and the early radiation in the Hayden-Preskill recovery problem respectively. As such, emergence of the volume-law entanglement in monitored quantum circuits can be interpreted as information recovery from an old black hole via projective measurements of outgoing radiations. 

Furthermore, the entanglement distillation algorithm can be converted into an algorithm to reconstruct the initial quantum information that was thrown into an old black hole. For the case of Clifford dynamics, the recovery algorithm is given by
\begin{align}
 \figbox{1.5}{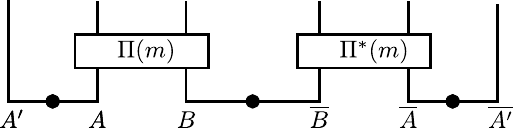} \ .
\end{align}
where the dual code will be constructed by reversing the flow of the time. After applying an appropriate feedback Pauli operator, EPR pairs will be distilled on $A'$ and $\overline{A'}$.

The physics of monitored quantum circuits also provides us with useful insights into the problem of the black hole interior reconstruction. Understanding the black hole interior will be essential in resolving various puzzles concerning the quantum nature of black holes. According to Hawking's semiclassical calculation, there must exist pairs of entangled Rindler modes across the black hole horizon. Then, the modes inside the black hole can then be defined unambiguously in the outside quantum mechanics language as an entangled partner of the outgoing mode. As such, explicitly writing down degrees of freedom that are entangled with the outgoing mode is an important issue.

A monitored quantum circuit can be interpreted as a toy model of the interior reconstruction problem where the black hole is continuously measured by outside observers~\cite{Beni19}. Namely, the subsystem $A$ can be viewed as the outgoing mode, and finding degrees of freedom, which are entangled with $A$, is equivalent to identifying the interior partner mode. Here $R$ can be viewed as the early Hawking radiation that was entangled with the black hole initially. When a black hole remains unperturbed with no projective measurement, the outgoing mode $A$ is entangled with the reference system $R$, which suggests that the interior partner mode can be found on $R$. When a black hole is continuously monitored, the outgoing mode $A$ will be decoupled from the reference system $R$, and will be entangled with the complementary subsystem $B$. This suggests that the interior partner mode can be written in a state-independent manner by using degrees of freedom on $B$ only. Namely, the same construction of the interior partner mode works for the cases where the black hole's initial state was a pure state. 

\section{Outlook}\label{sec:discussion}

In this paper, we have investigated the entanglement structure in monitored Clifford circuits and presented a method of verifying the entanglement. The main technical tool was the use of a dual classical error-correcting code whose codewords correspond to the spacetime patterns of the operator growth measured by OTOCs. We have also applied the developed framework to study the coding properties of monitored Clifford circuits. Finally, we have applied our technical results to various physical questions and puzzles. We hope that theoretical techniques developed in this paper will be useful in further addressing various important open problems concerning monitored Clifford circuits and beyond. Below we discuss some possible future problems. 

We have presented a simple deterministic entanglement distillation algorithm that enables us to verify quantum entanglement between two subsystems in a monitored Clifford circuit. We expect that this algorithm can be readily employed for experimental demonstrations of quantum entanglement arising in a monitored Clifford circuit. It is worth recalling that recently~\cite{Noel:aa2021} has reported an experimental demonstration of quantum error-correction properties (\emph{i.e.} entanglement between the reference system $R$ and the system) in a monitored Clifford circuit. Our main focus here is to directly verify quantum entanglement in the system without using the reference system. We have also pointed out that a monitored circuit problem is fundamentally akin (or actually identical) to the Hayden-Preskill recovery problem by reversing the flow of time. The Hayden-Preskill recovery algorithm has been experimentally demonstrated, see~\cite{Landsman:2019aa, Blok:2021aa} for instance. We expect that similar experimental setups can be utilized to verify the entanglement structure in a monitored quantum circuit.  

In this paper, we mainly focused on developing theoretical techniques to investigate the entanglement structure in monitored Clifford circuits without looking at specific models. The next step is to apply our framework to concrete models. A potentially interesting example is a random monitored Clifford circuit where both codeword and error vectors will have random entries which may give us some analytical control in computing coding properties. Also, we expect that our technique is useful in addressing the cases where the time evolution and measurements are translation symmetric in space and time. For such situations, polynomial representations of Pauli operators may be utilized~\cite{Haah13, Beni13}.

Another interesting future problem concerns the entanglement structure in generic monitored quantum circuits beyond Clifford dynamics. Naive applications of ideas from~\cite{Yoshida:2017aa}, or the Petz recovery map~\cite{Ohya_Petz_Text}, would lead to a distillation algorithm which post-selects the measurement results to satisfy $m=\overline{m}$ (or in other words, $s=\vec{1}$). Unfortunately, the success probability will be rather small, and turning it into a deterministic algorithm will increase the circuit complexity by a huge factor. In this paper, for Clifford circuits, we have found that our entanglement distillation algorithm succeeds even without any feedback as long as $s = m\cdot \overline{m} \in \E$. This leaves a hope that the post-selection probability may not be pessimistically small for generic monitored circuits as well.  Relatedly, we expect that modification of traversable wormhole protocols may provide efficient distillation methods~\cite{Brown:aa, Nezami:aa, Gao:aa, Schuster:aa}.

It is interesting to note that insertion of boundaries ($D$-branes) in the AdS/BCFT correspondence has an effect similar to projecting a subsystem onto a random state~\cite{Takayanagi:2011aa, Fujita:2011aa}. This suggests that insertion of boundaries may be interpreted as a projective measurement that realizes situations analogous to Eq.~\eqref{eq:D-Brane}. It has been suggested that the effect of placing an end-of-the-world (EoW) brane on a two-sided AdS black hole is the same as projecting a quantum state to a particular pure quantum state~\cite{Almheiri:aa}. It will be interesting to test this proposal further by using tensor network toy models~\cite{Pastawski15b, Hayden:2016aa}. It is also useful to note that an effective action for entanglement entropy of monitored quantum circuits advocated in~\cite{Li:2021aa} (capillary-wave theory) contains a term which can be viewed as surface tension. 

It is important to note that the volume-law entanglement exists because one records the measurement outcomes. If projective measurements were performed, but the measurement outcomes were forgotten, the total effect can be modelled as a dephasing channel. It will be interesting to consider the cases where measurement outcomes are partially forgotten. Our framework of mapping to a dual classical code may suggest a possibility that a little bit of forgetfulness can be tolerated when the encoding into codewords $\C(P_{A})$ is robust.

As is evident from the construction of error vectors, the causal ordering of projective measurements $P_{1},\cdots, P_{\tau}$ is crucial. Then, given a set of measured operators which are not necessarily sorted in a chronological order, it will be interesting to ask if the causal ordering (\emph{i.e.} the arrow of time) can be inferred from the output wavefunction or not. 

\subsection*{Acknowledgment}

I thank Tim Hsieh and Zhi Li for useful discussions. Research at the Perimeter Institute is supported by the Government of Canada through Innovation, Science and Economic Development Canada and by the Province of Ontario through the Ministry of Economic Development, Job Creation and Trade.

\appendix

\section{Measurement probability (Proof of lemma~\ref{lemma:sum})}\label{sec:probability}

In this section, we prove lemma~\ref{lemma:sum} by evaluating $\Prob(m,\overline{m})$.

\subsection{Measurement probability}

Let us compute the probability of measuring $m$ and $\overline{m}$. It is given graphically as follows:
\begin{align}
\text{Prob}(m,\overline{m}) = \  \figbox{1.5}{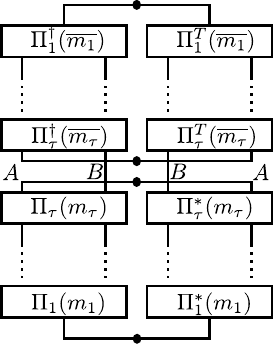}\ .
\end{align}
Here, in order to make the figure smaller, we moved some diagrams to the right hand side of the system. Specifically, we employed the following rule which applies to arbitrary operators:
\begin{align}
 \figbox{1.5}{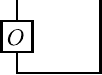} \ = \  \figbox{1.5}{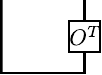}\ 
\end{align}
where $O^{T}$ represents a transpose of $O$.

One can rewrite the above expression as follows:
\begin{align}
\text{Prob}(m,\overline{m}) = \frac{1}{d_{A}^2} \sum_{P_{A}\in \text{Pauli}_{A}}\Big\langle  \Pi^{\dagger}(m) P_{A}^{\dagger} \Pi(\overline{m}) \Pi^{\dagger}(\overline{m}) P_{A} \Pi(m)  \Big\rangle 
\end{align}
where we inserted the summation over Pauli operators on $A$, namely
\begin{align}
\frac{1}{d_{A}}\sum_{P_{A}} P_{A}\otimes P_{\overline{A}}^{\dagger}=\text{SWAP}_{A\overline{A}}. 
\end{align}
Here it is convenient to introduce the following function:
\begin{equation}
\begin{split}
\Prob (m,\overline{m}; P_{A}) &\equiv 
\Big\langle  \Pi^{\dagger}(m) P_{A}^{\dagger} \Pi(\overline{m}) \Pi^{\dagger}(\overline{m}) P_{A} \Pi(m)  \Big\rangle 
=\ \figbox{1.5}{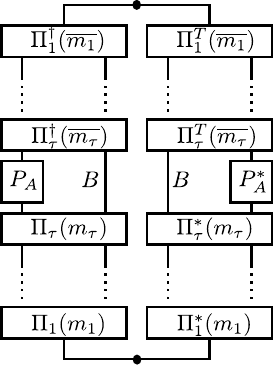}\ . \label{eq:Prob(m,m,PA)}
\end{split}
\end{equation}
Thus we arrived at the following lemma.

\begin{lemma}\label{lemma:average}
We have
\begin{align}
\Prob(m,\overline{m}) = \frac{1}{d_{A}^2} \sum_{P_{A}\in \Pauli_{A}} \Prob(m,\overline{m}; P_{A}).
\end{align}
\end{lemma}

In other words, $\Prob(m,\overline{m})$ is the average of $\Prob(m,\overline{m}; P_{A})$ taken over all the Pauli operators $P_{A}$ on $A$. 

The following lemma, concerning properties of $\Prob(m,\overline{m}; P_{A})$, will be useful.

\begin{lemma}\label{lemma:commutation}
We have 
\begin{align}
\Prob(m,\overline{m}; P_{A}) = \Prob(m,\overline{m} \cdot \C(P_{A}); I_{A}).
\end{align}
\end{lemma}

The proof of lemma~\ref{lemma:commutation} is immediate from the following observation:
\begin{align}
\Pi_{j}(m_{j}) P_{A} = P_{A}\Pi_{j}(m_{j}\cdot \C(P_{A})_{j} )
\end{align}
since
\begin{align}
P_{j} P_{A} = \C(P_{A})_{j}P_{A}P_{j} \qquad \C(P_{A})_{j} = \pm 1.
\end{align}
Hence, we have
\begin{equation}
\begin{split}
\Prob (m,\overline{m}; P_{A}) &= \Big\langle  \Pi^{\dagger}(m) P_{A}^{\dagger} \Pi(\overline{m}) \Pi^{\dagger}(\overline{m}) P_{A} \Pi(m)  \Big\rangle  \\
&= \Big\langle  \Pi^{\dagger}(m)  \Pi(\overline{m}\cdot \C(P_{A}))P_{A}^{\dagger}P_{A} \Pi^{\dagger}(\overline{m}\cdot \C(P_{A}))  \Pi(m)  \Big\rangle \\
&= \Big\langle  \Pi^{\dagger}(m)  \Pi(\overline{m}\cdot \C(P_{A})) \Pi^{\dagger}(\overline{m}\cdot \C(P_{A}))  \Pi(m)  \Big\rangle \\
&= \Prob(m,\overline{m} \cdot \C(P_{A}); I_{A}).
\end{split}
\end{equation}

\subsection{Summation of measurement probability}

As we mentioned earlier, our primary focus will be on $s = m\cdot \overline{m}$. Hence it is convenient to define the summation of probabilities over $m$ as follows:
\begin{align}
\Sum(s) \equiv \sum_{m}\Prob(m, m\cdot s ) \qquad \Sum(s; P_{A}) \equiv \sum_{m}\Prob(m, m\cdot s ; P_{A} ).
\end{align}
We can verify 
\begin{align}
\Sum(s) = \frac{1}{d_{A}^2}\sum_{P_{A}\in \Pauli_A} \Sum(s; P_{A}).
\end{align}

The central result of this section is the following lemma. 

\begin{lemma}\label{lemma:probability}
We have
\begin{equation}
\begin{split}
\Sum(s;P_{A}) \equiv \sum_{m} \Prob(m, m\cdot s; P_{A}) &=  \frac{1}{d_{\mathcal{E}}} \qquad s \in \mathcal{E}^{(P_{A})} \\ 
&= 0 \qquad \ \ s \not\in \E^{(P_{A})}
\end{split}
\end{equation}
where $d_{\mathcal{E}}$ is the number of elements in $\E^{(P_{A})}$.
\end{lemma}

Due to lemma~\ref{lemma:commutation}, it suffices to prove lemma~\ref{lemma:probability} for $P_{A}=I_{A}$, namely
\begin{align}
\Sum(s;I_{A}) = \sum_{m} \Prob(m, m\cdot s; I_{A}) =  \frac{1}{d_{\mathcal{E}}} \qquad s \in \mathcal{E}^{(I_A)}.
\end{align}
We will prove this statement in the next subsection.

With lemma~\ref{lemma:probability} in hand, one can easily prove lemma~\ref{lemma:sum}. Namely, we have 
\begin{align}
\Sum(s) = \frac{1}{d_{A}^2} \sum_{P_{A}\in \Pauli_A} \Sum(s;P_{A}).
\end{align}
Hence $\Sum(s)=0$ when $s\not\in \E_{\text{total}}$. For $s\in \E_{\text{total}}$, $\Sum(s)$ takes a uniform value. Hence we arrive at 
\begin{align}
\Sum(s) = \frac{1}{d_{\E_{\text{total}}}} \qquad s\in \E_{\text{total}}.
\end{align}
This completes the proof of lemma~\ref{lemma:sum}.

\subsection{Proof of lemma~\ref{lemma:probability}}

The proof of lemma~\ref{lemma:probability} proceeds by induction, so it is convenient to denote the lemma with ``$t$-index'':
\begin{align}
\Sum^{(t)}(s^{(t)};I_{A}) = \sum_{m^{(t)}} \Prob^{(t)}(m^{(t)}, m^{(t)}\cdot s^{(t)}; I_{A}) =  \frac{1}{d_{\mathcal{E}^{(t)}}} \qquad s \in \mathcal{E}^{(t)}
\end{align}
for a monitored Clifford circuit with $t$ measurements of $P_{1},\cdots, P_{t}$.

For $t = 1$, we have 
\begin{align}
\Prob^{(1)}(m_{1},\overline{m_{1}};I_{A}) = \ \figbox{1.5}{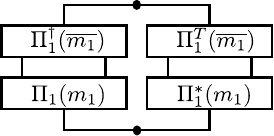}\   = \frac{1}{2} \delta_{m_{1},\overline{m_{1}}}
\end{align}
and
\begin{align}
\Sum^{(1)}(s_{1};I_{A}) = \sum_{m_{1}} \Prob(m_{1},\overline{m_{1}};I_{A}) = \delta_{s_{1},1}.
\end{align}
The error vector set is given by $\E^{(1)} = \{ (1)\}$ since $\E^{(1)}(P_{1})= (1)$. So, the lemma holds for $t=1$. 

Next, let us assume that the lemma holds for $t=\tau - 1$ and show that the lemma holds also for $t=\tau$. We have
\begin{align}
\Prob^{(\tau)}(m^{(\tau)},\overline{m}^{(\tau)};I_{A}) =  \ \figbox{1.5}{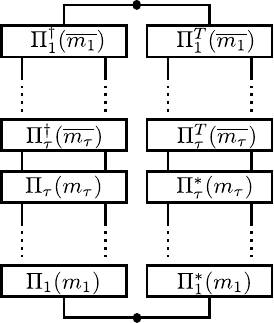}\ = \delta_{m_{\tau},\overline{m_{\tau}}} \ \figbox{1.5}{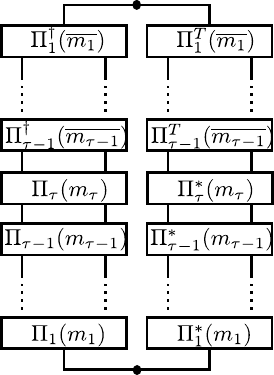}\ 
\end{align}
where $m^{(\tau)}=(m_{1},\cdots, m_{\tau})$ and $\overline{m}^{(\tau)}=(\overline{m_{1}},\cdots, \overline{m_{\tau}})$. Here we used $\Pi_{\tau}^{\dagger}(\overline{m_{\tau}})\Pi_{\tau}(m_{\tau})=\delta_{ m_{\tau},\overline{m_{\tau}}} \Pi_{\tau}(m_{\tau})$.

To compute $\Sum^{(\tau)}(s^{(\tau)};I_{A})$, we set $\overline{m}^{(\tau)}=m^{(\tau)}\cdot s^{(\tau)}$ and sum over $m^{(\tau)}$. Summing over $m_{\tau}$ gives
\begin{align}
\sum_{m_{\tau}} \Prob^{(\tau)}(m^{(\tau)},\overline{m}^{(\tau)};I_{A}) = \delta_{s_{\tau},1} \sum_{m_{\tau}}\ \figbox{1.5}{fig-Prob_m,m,IA_modified.pdf}\ .
\end{align}

Let us evaluate $\Pi_{\tau-1}^{\dagger}(\overline{m_{\tau - 1}}) \Pi_{\tau}(m_{\tau})\Pi_{\tau-1}(m_{\tau - 1})$. We have 
\begin{equation}
\begin{split}
\Pi_{\tau-1}^{\dagger}(\overline{m_{\tau - 1}}) \Pi_{\tau}(m_{\tau})\Pi_{\tau-1}(m_{\tau - 1}) &=\Pi_{\tau-1}^{\dagger}(\overline{m_{\tau - 1}})\left(\frac{I+m_{\tau}P_{\tau}}{2}\right) \Pi_{\tau-1}(m_{\tau - 1}) \\
&=\frac{1}{2}\Pi_{\tau-1}^{\dagger}(\overline{m_{\tau - 1}}) \Pi_{\tau-1}(m_{\tau - 1}) + \frac{m_{1}}{2}\Pi_{\tau-1}(\overline{m_{\tau - 1}})^{\dagger}  P_{\tau} \Pi_{\tau-1}(m_{\tau - 1}).
\end{split}
\end{equation}
Since $\Pi_{\tau-1}^{\dagger}(\overline{m_{\tau - 1}}) \Pi_{\tau}(m_{\tau})\Pi_{\tau-1}(m_{\tau - 1})$ appears twice in the expression of $\Prob^{(\tau)}(m^{(\tau)},\overline{m}^{(\tau)};I_{A})$, this decomposition generates four terms. Terms linear in $m_{\tau}$ will vanish when we take sum over $m_{\tau}$. Hence we have 
\begin{equation}
\begin{split}
\sum_{m_{\tau}} \Prob^{(\tau)}(m^{(\tau)},\overline{m}^{(\tau)};I_{A}) &= \frac{1}{2}\delta_{s_{\tau},1} \left( \ \figbox{1.5}{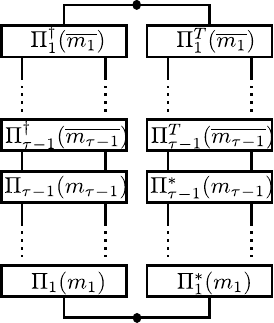}\ + \ \figbox{1.5}{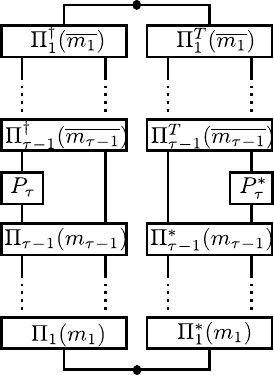} \ \right)\ \\
&=\frac{1}{2}\delta_{s_{\tau},1} \left( \Prob^{(\tau-1)}(m^{(\tau)},\overline{m}^{(\tau)};I_{A})  + \Prob^{(\tau-1)}(m^{(\tau-1)},\overline{m}^{(\tau-1)}\cdot \E(P_{\tau})^{(\tau-1)};I_{A})  \right).
\end{split}
\end{equation}
Here we observed that the first diagram is identical to $\Prob^{(\tau-1)}(m^{(\tau-1)},\overline{m}^{(\tau-1)};I_{A})$ with $m^{(\tau-1)}=(m_{1},\cdots, m_{\tau-1})$ and $\overline{m}^{(\tau-1)}=(\overline{m_{1}},\cdots, \overline{m_{\tau-1}})$. As for the second diagram, commuting $P_{\tau}$ through and eliminating two $P_{\tau}$'s change $\overline{m_{j}}$ as follows
\begin{align}
\overline{m_{j}} \ \longrightarrow \ \overline{m_{j}}\cdot \E(P_{\tau})_{j} \qquad j=1,\cdots, \tau -1.
\end{align}
So, the second diagram is identical to $\Prob^{(\tau-1)}\big(m^{(\tau-1)},\overline{m}^{(\tau-1)}\cdot \E(P_{\tau})^{(\tau-1)};I_{A}\big)$. Here we defined $\E(P_{\tau})^{(\tau-1)}$ as a $\tau-1$-component vector:
\begin{align}
\E(P_{\tau})^{(\tau-1)}=\big(\E(P_{\tau})_{1},\cdots, \E(P_{\tau})_{\tau-1}\big)
\end{align}
by removing the $\tau$-th component $\E(P_{\tau})_{\tau}$. 

By taking summation over $m_{1},\cdots, m_{\tau-1}$, we have
\begin{align}
\Sum^{(\tau)}(s^{(\tau)};I_{A}) = \frac{1}{2}\delta_{s_{\tau},1} \Big( 
\Sum^{(\tau-1)}(s^{(\tau-1)};I_{A})  + \Sum^{(\tau-1)}(s^{(\tau-1)}\cdot \E(P_{\tau})^{(\tau-1)};I_{A}) \Big). \label{eq:final_summation}
\end{align}
By using the lemma for $t=\tau-1$, we have
\begin{align}
\Sum^{(\tau-1)}(s^{(\tau-1)};I_{A})=\frac{1}{d_{\E^{(\tau - 1)}}} \qquad s^{(\tau-1)}\in \E^{(\tau-1)}
\end{align}
and 
\begin{align}
\Sum^{(\tau-1)}(s^{(\tau-1)}\cdot \E(P_{\tau})^{(\tau-1)};I_{A}) = \frac{1}{d_{\E^{(\tau - 1)}}} \qquad s^{(\tau-1)}\cdot \E(P_{\tau})^{(\tau-1)} \in \E^{(\tau-1)}.
\end{align}

The remaining task is to explicitly compute Eq.~\eqref{eq:final_summation}. It is convenient to consider two cases separately. 

\begin{itemize}
\item If $\E(P_{\tau})^{(\tau-1)} \in \E^{(\tau-1)}$, we have
\begin{align}
d_{ \E^{(\tau-1)} }=d_{\E^{(\tau)}}
\end{align}
and
\begin{align}
\Sum^{(\tau)}(s^{(\tau)};I_{A}) =\delta_{s_{\tau},1} 
\Sum^{(\tau-1)}(s^{(\tau-1)};I_{A}) =  \delta_{s_{\tau},1}\frac{1}{d_{\E^{(\tau-1)}}}= \delta_{s_{\tau},1}\frac{1}{d_{\E^{(\tau)}}} \qquad s^{(\tau-1)}\in \E^{(\tau-1)}. 
\end{align}
Note that $\Sum^{(\tau)}(s^{(\tau)};I_{A})$ is nonzero only when $s_{\tau}=1$, \emph{i.e.} 
\begin{align}
s^{(\tau)} = \big( s_{1},\cdots, s_{\tau-1}, 1 \big) \qquad s^{(\tau-1)}\in \E^{(\tau-1)}.
\end{align}
This condition is equivalent to
\begin{align}
s^{(\tau)}\in \E^{(\tau)}
\end{align}
since $\E(P_{\tau})^{(\tau-1)} \in \E^{(\tau-1)}$. Hence, the lemma holds. 

\item If $\E(P_{\tau})^{(\tau-1)} \not\in \E^{(\tau-1)}$, we have 
\begin{align}
2d_{ \E^{(\tau-1)} }=d_{\E^{(\tau)}}
\end{align}
and 
\begin{align}
\Sum^{(\tau)}(s^{(\tau)};I_{A}) =\frac{1}{2}\delta_{s_{\tau},1}  \frac{1}{d_{ \E^{(\tau-1)} }} = \delta_{s_{\tau},1}  \frac{1}{d_{ \E^{(\tau)} }} 
\qquad s^{(\tau-1)}\in \E^{(\tau-1)}\ \ \text{or} \ \ s^{(\tau-1)}\in\E(P_{\tau})^{(\tau-1)}\cdot \E^{(\tau-1)}.
\end{align}
The condition for nonzero $\Sum^{(\tau)}(s^{(\tau)};I_{A})$ is equivalent to 
\begin{align}
s^{\tau} \in \E^{(\tau)},
\end{align}
so, the lemma holds. 

\end{itemize}

Hence, we have proved that the lemma holds for $t=\tau$ as well. This completes the proof of lemma~\ref{lemma:probability} for arbitrary $t$ by induction.

\section{Output of distillation algorithm (Proof of lemma~\ref{lemma:output})}\label{sec:output}

In this section, we show that the aforementioned distillation algorithm outputs EPR pairs on $A\overline{A}$ when the classical error-correction condition is satisfied. 

\subsection{No feedback}

We begin by discussing the cases where the measurement result satisfies $s=m\cdot \overline{m} \in \E^{(I_{A})}$. For these cases, no feedback operation is needed. We explicitly find that the EPR fidelity is unity. 

When the measurement outcome is $m$ and $\overline{m}$, the EPR fidelity (an overlap with $|\text{EPR}\rangle_{A\overline{A}}$) is given by 
\begin{equation}
\begin{split}
F^{|\text{EPR}\rangle_{A\overline{A}}}(m,\overline{m}) =\frac{1}{d_{A}^2}\frac{\Prob(m,\overline{m};I_{A})}{\Prob(m,\overline{m})}   &\not = 0 \qquad s \in \E^{(I_{A})} \\
&= 0 \qquad s \not\in \E^{(I_{A})}
\end{split}
\end{equation}
where the factor of $\Prob(m,\overline{m})^{-1}$ comes from the normalization of the wavefunction. From lemma~\ref{lemma:probability}, we see that the fidelity is nonzero only when $s\in \E^{(I_{A})}$. So, the cases with $s\not\in \E^{(I_{A})}$ will require some feedback operations.

The probability of measuring $s\in \E^{(I_{A})}$ is given by
\begin{align}
\sum_{s\in \E^{(I_{A})}}\sum_{m} \Prob(m, m\cdot s) = \sum_{s\in \E^{(I_{A})}} \Sum(s).
\end{align}
By post-selecting the measurement outcome to be $s\in \E^{(I_{A})}$, the probability of having $m,\overline{m}$ with $s\in \E^{(I_{A})}$ is given by
\begin{align}
\frac{\Prob(m,\overline{m})}{\sum_{s\in \E^{(I_{A})}} \Sum(s)} \qquad s\in \E^{(I_{A})}.
\end{align}
The average fidelity is given by
\begin{equation}
\begin{split}
\text{Fidelity} &= \sum_{s\in \E^{(I_{A})}}\sum_{m} \frac{\Prob(m,\overline{m})}{\sum_{s\in \E^{(I_{A})}} \Sum(s)} F^{|\text{EPR}\rangle_{A\overline{A}}}(m,\overline{m}) \\
&= \frac{1}{d_{A}^2}\frac{  \sum_{s\in \E^{(I_{A})}}\sum_{m}  \Prob(m,\overline{m};I_{A}) }{ \sum_{s\in \E^{(I_{A})}} \Sum(s) } \\
&= \frac{  \sum_{s\in \E^{(I_{A})}} \Sum(s;I_{A})  }{ \sum_{P_{A}\in \Pauli_{A}}\sum_{s\in \E^{(I_{A})}} \Sum(s;P_{A}) } \\
&=1
\end{split}
\end{equation}
Here we used the error-correction condition. Namely, due to lemma~\ref{lemma:probability}, for $P_{A}\not=I_{A}$, we have
\begin{align}
\Sum(s;P_{A}) = 0 \qquad s \in \E^{(I_{A})}.
\end{align}

\subsection{With feedback}

Next, let us consider the cases where the measurement result satisfies $s=m\cdot \overline{m} \not\in \E^{(I_{A})}$. For these cases, we show that the output state, before applying the feedback, is given by the Choi-Jamio\l{}kowski state $|P_{A}\rangle \equiv (P_{A}\otimes I_{\overline{A}}) |\text{EPR}\rangle_{A\overline{A}}$.
The fidelity for $|P_{A}\rangle$ is given by 
\begin{equation}
\begin{split}
F^{|P_{A}\rangle}(m,\overline{m}) =\frac{1}{d_{A}^2} \frac{\Prob(m,\overline{m};P_{A}) }{\Prob(m,\overline{m})} &\not = 0 \qquad s \in \E^{(P_{A})} \\
& = 0 \qquad s \not \in \E^{(P_{A})}.
\end{split}
\end{equation}
By using lemma~\ref{lemma:commutation}, we have
\begin{align}
F^{|P_{A}\rangle}(m,\overline{m}) =\frac{1}{d_{A}^2} \frac{\Prob(m,\overline{m}\cdot \C(P_{A});I_{A})   }{\Prob(m,\overline{m})}. \label{eq:feedback}
\end{align}
For $s \in \E^{(P_{A})}$, we have
\begin{align}
m \cdot \overline{m} \cdot \C(P_{A}) \in \E^{(I_{A})}.
\end{align}
So, the calculation of Eq.~\eqref{eq:feedback} can be reduced to the case with $s\in \E^{(I_{A})}$. Hence the average fidelity for $|P_{A}\rangle$ is unity. By applying the feedback Pauli operator $P_{A}$, we obtain $|I_{A}\rangle = |\text{EPR}\rangle$. Therefore, the aforementioned distillation algorithm outputs EPR pairs deterministically if the classical error-correction condition is satisfied. 

\subsection{Imperfect cases}

We have shown that the distillation algorithm outputs EPR pairs deterministically when the classical error-correction condition is satisfied. Finally, we compute the output from the distillation algorithm when the condition is not satisfied. 

Since the feedback operation effectively reduces the problem to the cases with $s\in \E^{(I_{A})}$, it suffices to compute the output quantum state for $s\in \E^{(I_{A})}$. We will explicitly decompose the output wavefunction by using the Choi-Jamio\l{}kowski state of $P_{A}$. Namely, we will compute the overlap with $|P_{A}\rangle \langle Q_{A}|$ for $P_{A},Q_A \in \Pauli_A$. We have 
\begin{align}
F^{|P_{A}\rangle \langle Q_{A}|} (m,\overline{m}) = \frac{1}{\Prob(m,\overline{m})} \frac{1}{d_{A}^2} \ \figbox{1.5}{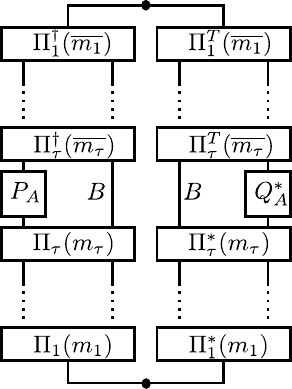}.
\end{align}

We begin with the cases with $P_{A}=Q_{A}$. We have 
\begin{equation}
\begin{split}
F^{|P_{A}\rangle \langle P_{A}|} (m,\overline{m}) &=  \frac{1}{d_{A}^2}  \frac{\Prob(m,\overline{m};P_{A})}{\Prob(m,\overline{m})} \qquad (s \in \E^{(I_{A})}),
\end{split}
\end{equation}
so the averaged fidelity is 
\begin{align}
\text{Fidelity}^{|P_{A}\rangle \langle P_{A}|} =  \frac{1}{d_{A}^2} \frac{\sum_{s\in \E^{(I_{A})} }\Sum(s;P_{A})}{\sum_{s\in \E^{(I_{A})}} \Sum(s)} =  \frac{\sum_{s\in \E^{(I_{A})} }\Sum(s;P_{A})}{\sum_{P_{A}\in \Pauli_A}\sum_{s\in \E^{(I_{A})}} \Sum(s;P_{A})}.
\end{align}
Hence we have 
\begin{equation}
\begin{split}
\text{Fidelity}^{|P_{A}\rangle \langle P_{A}|}  &= \frac{1}{N_{I_{A}}} \qquad \E^{(P_{A})} =  \E^{(I_{A})}  \\
&= 0 \qquad \ \quad \E^{(P_{A})} \not=  \E^{(I_{A})}.
\end{split}
\end{equation}
where $N_{I_{A}}$ is the number of Pauli operators $Q_{A}$ such that $\E^{(Q_{A})}=\E^{(I_{A})}$. 

Next, we study the cases where $P_{A}\not=I_{A}$ and $Q_{A}=I_{A}$. The central object to study is the following diagram:
\begin{align}
\figbox{1.5}{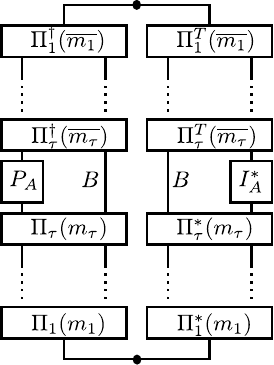}
\end{align}
Recall that $\Tr(P_{A})=0$ when $P_{A}\not=I$. Also recall that $\Pi_{j}(m_{j})$ consists of an identity operator and a Pauli operator $m_{j}P_{j}$. So, in order to have a non-trivial contribution, some combinations of $P_{j}$'s in $\Pi(m), \Pi(\overline{m})$ need to generate $P_{A}$. In other words, there must exist a set of indices $\Lambda \in \{1,\cdots, \tau\}$ such that
\begin{align}
P_{A} \propto \prod_{j\in \Lambda} P_{j}
\end{align}
up to a $U(1)$ phase. In the above diagram, $P_{j}$ appear four times. In order to generate a term proportional to $P_{A}$, $P_{j}$ for $j\in \Lambda$ needs to be multiplied odd times because $P_{j}^2=I$. Hence, possible coefficients of $P_{j}$ are 
\begin{align}
m_{j}, \ \overline{m_{j}}, \ m_{j}^2 \overline{m_{j}}, \ m_{j} \overline{m_{j}}^2.
\end{align}
By fixing $s\in \E^{(I_{A})}$, let us take summation over $m$. We see that terms with coefficients $m_{j}$ and $m_{j} \overline{m_{j}}^2$ vanish. Also, since $\overline{m_{j}}=s_{j}\cdot m_{j}$, terms with coefficients $\overline{m_{j}}$ and $m_{j}^2 \overline{m_{j}}$ vanish. Hence we arrive at 
\begin{align}
\sum_{s\in \E^{(I_{A})}} \sum_{m}\ \figbox{1.5}{fig-Prob_m,m,IA_induction.pdf}\ = 0 \qquad (P_{A}\not=I_{A}).
\end{align}
So, there is no contribution to $|P_{A}\rangle \langle I_{A}|$ when $P_{A}\not=I_{A}$. 

Finally, let us study the cases where $P_{A}\not=Q_{A}$. Since commuting $Q_{A}$ through change $\overline{m}$ to $\overline{m}\cdot \C(Q_{A})$ and $P_{A}$ to $P_{A}Q_{A}^\dagger$, the analysis from the previous paragraph (with $P_{A}\not=I_{A}$ and $Q_{A}=I_{A}$) can be applied. So, one can conclude that there is no contribution to $|P_{A}\rangle \langle Q_{A}|$ with $P_{A}\not=Q_{A}$. Hence, we arrive at 
\begin{align}
\rho_{A\overline{A}} = \frac{1}{N_{I_{A}}}\sum_{P_{A}:\E^{(P_{A})}=\E^{(I_{A})}} |P_{A}\rangle \langle P_{A}|. 
\end{align}
This proves lemma~\ref{lemma:output}.

\section{Conditional entropy (Proof of theorem~\ref{theorem:coherent})}\label{sec:entropy}

In this section, we will compute the conditional entropy $S_{A|B}\equiv S_{AB}-S_{B}$. 

The output of the monitored circuit is given by Eq.~\eqref{eq:output_monitor}, which is reprinted below:
\begin{align}
|\Psi(m)\rangle = \frac{1}{\sqrt{\Prob(m)}} \  \figbox{1.5}{fig-MQC3.pdf} \ .
\end{align}
For an output of a Clifford circuit, it suffices to compute the R\'{e}nyi-$2$ entropies. So, we have 
\begin{align}
2^{S_{A|B}(m)} =  2^{S_{AB}^{(2)}(m) - S_{B}^{(2)}(m) } = \frac{\Tr \big[ \rho_{B}(m)^2 \big] }{\Tr \big[ \rho_{AB}(m)^2 \big] }.
\end{align}

One can compute $\Tr\big[\rho_{B}(m)^2\big]$ by looking at $\Prob(m,m)$ (with $\overline{m}=m$). Namely we have the following relation:
\begin{align}
\Prob(m,m) =  \Tr \big[ \rho_{B}(m)^2 \big] \Prob(m)^2  \frac{d}{d_{A}}.
\end{align}
As for, $\Tr\big[\rho_{AB}(m)^2\big]$, we have 
\begin{align}
\Prob(m,m; I_{A}) =  \Tr \big[ \rho_{AB}(m)^2 \big] \Prob(m)^2  d.
\end{align}
Hence, we have 
\begin{align}
2^{S_{A|B}(m)} = d_{A} \frac{\Prob(m,m) }{\Prob(m,m;I_{A}) }.
\end{align}
This equality holds as long as $\Prob(m)\not=0$.

We have obtained an expression of the conditional entropy $S_{A|B}(m)$ for each realization $m$. It turns out that $S_{A|B}(m)$ does not depend on $m$. This results from the following fact:
\begin{itemize}
\item Both $ \Prob(m,m; I_{A})$ and $\Prob(m,m)$ do not depend on $m$ (as long as $\Prob(m)\not=0$).
\end{itemize}

This statement can be proven by using a certain property of Clifford circuits and stabilizer states. Here we sketch the proof idea. Recall that one can simulate the monitored Clifford circuit (consisting of Clifford gates with Pauli measurements) as a unitary Clifford circuit by adding ancilla qubits that record measurement results and entangling the circuit with ancilla qubits by Control-Not gates (which are Clifford operators). For instance, the output from the monitored circuit with ancilla qubits can be written as 
\begin{align}
\sum_{m} \sqrt{\Prob(m)}|\Psi(m)\rangle \otimes |m\rangle
\end{align}
where $|m\rangle=|m_{1},\cdots, m_{\tau}\rangle $ is defined on $\tau$ ancilla qubits. Note that the above quantum state is a stabilizer state, so its spectrum of reduced density matrices in subsystems must be flat (which can be proven in a standard manner, see~\cite{Fattal04} for instance). Looking at the reduced density matrix on the ancilla Hilbert space, its coefficient for $|m\rangle \langle m |$ corresponds to $\Prob(m)$ which must be uniform as long as $\Prob(m)\not=0$:
\begin{align}
\Prob(m)=\text{const} \qquad \text{if} \qquad \Prob(m)\not=0.
\end{align}
Similarly, one can construct stabilizer states whose reduced density matrices encode $ \Prob(m,m; I_{A})$ and $\Prob(m,m)$ as coefficients. Then we can prove that $ \Prob(m,m; I_{A})$ and $\Prob(m,m)$ are uniform and do not depend on $m$. 

By using this property of Clifford quantum circuits, we arrive at 
\begin{equation}
\begin{split}
2^{S_{A|B}(m)} &= d_{A} \frac{\Sum(s=1)}{\Sum(s=1;I_{A})} \\
&= \frac{1}{d_{A}} \frac{\sum_{P_{A}\in \Pauli_A } \Sum(s=1;P_{A})}{\Sum(s=1;I_{A})}  \\
&= \frac{N_{I_{A}}}{d_{A}}.
\end{split}
\end{equation}
This completes the proof of theorem~\ref{theorem:coherent}). 

\section{Logical operators (Proof of lemma~\ref{lemma:logical})}\label{sec:system-proof}

In this section, we will prove
\begin{align}
\C(P) \in \E^{(I)} \quad \Leftrightarrow \quad P \in \mathcal{L}. \label{eq:logic}
\end{align}
The proof proceeds by induction. It is immediate to show that Eq.~\eqref{eq:logic} holds for $\tau = 1$ as $\Logic^{(1)}$ is defined as the commutant of $P_{1}$.  Here we assume that Eq.~\eqref{eq:logic} holds for $\tau -1$ and prove it for $\tau$.

Proof of $\Leftarrow$: Let us assume $P\in \Logic^{(\tau)}$ and prove $\C(P)\in \E^{(I)}$. It is useful to recall that if $P\in \Logic^{(\tau)}$, then $[P,P_{\tau}]=0$ since $P_{\tau}\in \Stab^{(\tau)}$.

\begin{itemize}
\item Case 1: $P\in \Logic^{(\tau - 1)}$.

In this case, we have $[P,P_{\tau}]=[P,P_{\tau-1}]=0$, so $\C(P)$'s entries for the $\tau$-th and $\tau-1$-th components are trivial. By using Eq.~\eqref{eq:logic} for $\tau - 1$, we see that there exists a set of indices $\Lambda \subseteq \{1,\cdots, \tau - 1 \}$ such that 
\begin{align}
\C(P) = \prod_{j\in \Lambda} \E(P_{j}) \in \E^{(I)}.
\end{align}

\item Case 2: $P\not\in \Logic^{(\tau - 1)}$. 

In this case, with some work, one can prove $PP_{\tau}\in \Logic^{(\tau-1)} $~\footnote{
Recall that $P\in \Logic^{(\tau )}$ implies that $P$ commutes with all the elements in $\Stab^{(\tau)}$ since $\Logic^{(\tau )}$ is the commutant of $\Stab^{(\tau)}$. But $P\not\in \Logic^{(\tau - 1)}$ implies that $P$ does not commute with some elements in $\Stab^{(\tau-1)}$. Here we can write independent stabilizer generators as $\Stab^{(\tau-1)} = \langle S_{1},S_{2},\cdots \rangle$ such that $[P,S_{1}]\not=0$ and $[P,S_{j}]=0$ for $j\geq 2$. 

Now, we prove $[P_{\tau},S_{1}]\not=0$ and $[P_{\tau},S_{j}]=0$ for $j\geq 2$. Let us begin with $[P_{\tau},S_{1}]\not=0$. Suppose $[P_{\tau}, S_{1}]=0$. Then, from the recursive construction of $\Stab^{(\tau)}$, we see that $S_{1} \in \Stab^{(\tau)}$. This contradicts with the fact that $[P,S_{1}]\not=0$, but $[P,\Stab^{(\tau)}]=0$. So, we have $[P_{\tau},S_{1}]\not=0$ . 

As for $[P_{\tau},S_{j}]=0$ for $j\geq 2$, let us focus on $j=2$. Suppose $[P_{\tau},S_{2}]\not=0$. Then we have $[P_{\tau},S_{1}S_{2}]=0$ which implies $S_{1}S_{2}\in \Stab^{(\tau)}$. But this contradicts with the fact that $[P,S_{1}]\not=0$ and $[P,S_{2}]=0$, but $P$ commutes with all the elements in $\Stab^{(\tau)}$. So we have $[P_{\tau},S_{j}]=0$ for $j\geq 2$.

From these arguments, one can show that $PP_{\tau}\in \Logic^{(\tau-1)}$.
}.

Since $PP_{\tau}\in \Logic^{(\tau)}$, one can apply the argument from Case 1 to $PP_{\tau}$. This shows 
\begin{align}
\C  (PP_{\tau}) \in \E^{(I)}.
\end{align}
Since $\C(P_{\tau}) \in \E^{(I)}$, we have $\C(P) \in \E^{(I)}$.
\end{itemize}

Proof of $\Rightarrow$: Let us assume $\C(P)\in \E^{(I)}$ and prove $P\in \Logic^{(\tau)}$. From this assumption, there exists a set of indices $\Lambda \in \{ 1,\cdots, \tau \}$ such that 
\begin{align}
\C(P) = \prod_{j \in \Lambda} \E(P_{j}). \label{eq:some_relation}
\end{align}

\begin{itemize}
\item Case 1: $\tau \not\in \Lambda$.

In this case, by using Eq.~\eqref{eq:logic} for $\tau - 1$, we see that $P\in\Logic^{(\tau-1)}$. Eq.~\eqref{eq:some_relation} implies that the $\tau$-th component of $\C(P)$ is trivial, and hence $[P,P_{\tau}]=0$. This implies 
\begin{align}
P \in \Logic^{(\tau)}.
\end{align}

\item Case 2: $\tau \in \Lambda$. 

In this case, we observe
\begin{align}
\C(PP_{\tau}) = \prod_{j \in \Lambda_{/ \tau} } \E(P_{j}). 
\end{align}
where $\Lambda_{/ \tau}$ means that $\tau$ is removed from $\Lambda$. Then, one can apply the argument from Case 1 to $PP_{\tau}$. This completes the proof.
\end{itemize}

\section{More on coding properties}\label{app:code}

In this appendix, we present additional results on the coding properties of a monitored Clifford circuit as well as proofs of some technical results.

\subsection{Extended codewords}

In this subsection, we will present an alternative derivation of the entanglement structure by treating the reference system $R$ as a part of the system. 

Instead of treating $AB$ as a system, we think of $ABR$ as the whole system. Specifically, imagine that there were initially $2n$ qubits in maximally mixed states and we perform Bell measurements in the following $2n$ Bell operators:
\begin{align}
X_{j}^{AB} \otimes X_{j}^{R} \qquad Z_{j}^{AB} \otimes Z_{j}^{R}  \qquad j=1,\cdots, n.
\end{align}
This will create a maximally entangled state on $ABR$. We then proceed to perform projective measurements $\Pi(m)$ on $A$. In this interpretation, the reference $R$ becomes a part of the system and we have a $2n$-qubit monitored quantum circuit where original $P_{j}$ measurements, as well as Bell measurements, are performed:
\begin{align}
 \figbox{1.5}{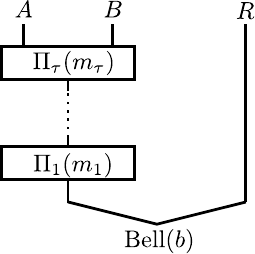} \ 
\end{align}
where $\Bell(b)$ represent Bell measurements with outcomes $b$. Starting from EPR pairs between $AB$ and $R$ is equivalent to postselecting the measurement outcomes to satisfy $b=(1,\cdots,1)$. 

In total, $\tau + 2n$ measurements are performed. Here it is convenient to define extended codeword and error vectors which include Bell measurements. Namely, we introduce an extended measurement vector
\begin{align}
m_{\ext} \equiv (m, b)
\end{align}
which has $\tau + 2n$ components. We can also define extended codeword and error vectors by $\C_{\ext}(P_{A})$ and $\E_{\ext}(P_{j})$ that account for commutation relations with respect to $2n$ Bell measurement operators as well as the original projective measurements of $P_{j}$. Extended error sets are denoted by $\E_{\ext}^{(P_{A})}$ with the error set $\E_{\ext}=\E_{\ext}^{(I_{A})}$. 

One can compute entanglement entropies of subsystems in terms of these extended vectors. By choosing Pauli operators on $A$, $B$ and $AB$ as initial information of the extended classical error-correcting code, we obtain the following three relations:
\begin{equation}
\begin{split}
S_{A|BR} &= - S_{A} = \log N_{{I_{A}}_{\ext}} - n_{A}\  \qquad N_{{I_{A}}_{\ext }} : \text{number of $P_{A}$ s.t. $\C_{\ext}(P_{A})\in \E_{\ext} $.} \\
S_{B|AR} &= - S_{B} = \log N_{{I_{B}}_{\ext}} - n_{B}  \qquad N_{{I_{B}}_{\ext }} : \text{number of $P_{B}$ s.t. $\C_{\ext}(P_{B})\in \E_{\ext} $.}\\
S_{AB|R} &= - S_{R} = \log N_{I_{\ext}} - n  \quad \ \qquad N_{I_{\ext }} : \text{number of $P$ s.t. $\C_{\ext}(P)\in \E_{\ext} $.} \label{eq:extended-entropy}
\end{split}
\end{equation}
It will be convenient to define the following three sets of Pauli operators:
\begin{equation}
\begin{split}
\mathcal{S} &\equiv \{ P \in \Pauli : \C(P) \in \E  \} \\
\mathcal{S}_{A} &\equiv \{ P_{A} \in \Pauli_{A} : \C(P_{A}) \in \E  \} \\
\mathcal{S}_{B} &\equiv \{ P_{B} \in \Pauli_{B} : \C(P_{B}) \in \E  \} 
\end{split}.
\end{equation}
Expert readers will recognize that Eq.~\eqref{eq:extended-entropy} is identical to the well-known formula for the entanglement entropy for a stabilizer state~\cite{Fattal04}:
\begin{align}
S_{R} = n_{R} - \log |\mathcal{S}_{R}|
\end{align}
where $R$ is an arbitrary subsystem and $\mathcal{S}_{R}$ is the restriction of the stabilizer group $\mathcal{S}$ onto $R$. Later, we will indeed show that these Pauli operators in $\mathcal{S}$ serve as stabilizer generators when the monitored Clifford circuit is viewed as a quantum error-correcting code. Using these relations, we obtain
\begin{equation}
\begin{split}
I_{(A,B)} &= \log \frac{N_{I_{\ext}}}{N_{{I_{A}}_{\ext}} N_{{I_{B}}_{\ext}} }\\
S_{AB|R} &= \log_{2} {N_{I}}_{\ext} - n.
\end{split}
\end{equation}

\subsection{Stabilizer operator from extended code}

The following lemma can be proven in a way similar to lemma~\ref{lemma:logical}. 

\begin{lemma}\label{lemma:stab}
Null operators in the extended code are stabilizer operators. Namely we have
\begin{equation}
\C_{\ext}(P) \in \E_{\ext} \qquad \text{\emph{iff}}\qquad P \in \mathcal{S}.
\end{equation}
\end{lemma}

\begin{figure}
\centering
\includegraphics[width=0.5\textwidth]{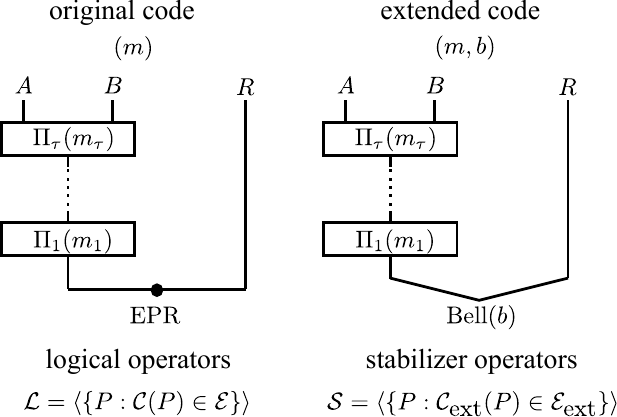} 
\caption{The original dual code and the extended dual code. The extended dual code is constructed by viewing $ABR$, including the reference $R$, as a whole system. Null vectors in two codes become logical operators and stabilizer operators. 
}
\label{fig-extended-idea}
\end{figure}

We will skip the proof. Our findings so far are summarized in Fig.~\ref{fig-extended-idea}.

\subsection{Cleaning lemma for monitored circuit}

We presented two different derivations of entanglement entropies by using the original and extended codewords respectively. This suggests that the numbers of Pauli operators in the logical operator groups ($N_{I}, N_{I_{A}}, N_{I_{B}}$) and those in the stabilizer group ($N_{I_{\ext}}, N_{{I_{A}}_{\ext}}, N_{{I_{B}}_{\ext}}$) can be related. Here we find the following three independent constraints:
\begin{align}
& \log N_{I} + \log N_{I_{A}} + \log N_{I_{B}}= \log N_{I_{\ext}} + \log N_{{I_{A}}_{\ext}} + \log N_{{I_{B}}_{\ext}} \label{eq:cleaning} \\ 
&\log N_{I_{A}} = 2n_{A} - \log N_{I_{\ext}} +\log  N_{{I_{B}}_{\ext}} \label{eq:cleaningA}  \\
&\log  N_{I_{B}} = 2n_{B} - \log N_{I_{\ext}} +\log  N_{{I_{A}}_{\ext}} \label{eq:cleaningB}  .
\end{align}
Below, we present coding theoretic interpretations of these equations. 

Let us begin with Eq.~\eqref{eq:cleaning}. Recall that $\log N_{I_{\ext}}$ is the number of independent stabilizer generators in $\mathcal{S}$. Also observe that $\log N_{I}$ is the number of independent stabilizer generators as well as independent logical operators. So, we have 
\begin{align}
g \equiv \log N_{I} - \log N_{I_{\ext}} =  \text{number of independent logical operators}.
\end{align}
Similarly, we find 
\begin{equation}
\begin{split}
&g_{A} \equiv \log N_{I_{A}} - \log N_{{I_{A}}_{\ext}} =  \text{number of independent logical operators supported on $A$} \\
&g_{B} \equiv \log N_{I_{B}} - \log N_{{I_{B}}_{\ext}} =  \text{number of independent logical operators supported on $B$}.
\end{split}
\end{equation}
With these interpretations, Eq.~\eqref{eq:cleaning} can be rewritten as
\begin{align}
g_{A} + g_{B} = g.
\end{align}
Noting that $g = 2k$ where $k$ is the number of logical qubits, this equation is identical to the cleaning lemma from~\cite{Beni10}.

Next, we look at Eq.~\eqref{eq:cleaningA}. Let us compute $N_{I_{A}}$ (the number of elements in $\mathcal{L}_{A}$) explicitly. Recall that logical operators on $A$ commute with all the stabilizer generators. Since they are supported exclusively on $A$, it suffices to look at independent stabilizer generators which have supports on $A$. In total, there are $\log N_{I_{\ext}} - \log N_{{I_{B}}_{\ext}}$ independent stabilizer generators with non-trivial supports on $A$. Recalling that there are $2n_{A}$ independent Pauli operators on $A$, we find
\begin{align}
\log N_{I_{A}} = 2n_{A} -  ( \log N_{I_{\ext}} - \log N_{{I_{B}}_{\ext}})
\end{align}
which is identical to Eq.~\eqref{eq:cleaningA}. Hence, Eq.~\eqref{eq:cleaningA} can be interpreted as a formula to compute the number of logical operators supported on $A$. Eq.~\eqref{eq:cleaningB} has a similar interpretation for logical operators on $B$.

Another interesting relation, which can be derived from the aforementioned three relations, is 
\begin{align}
\log N_{I} + \log N_{I_{\ext}} = 2n.
\end{align}
This follows from the fact that the logical operator group $\mathcal{L}$ is the commutant of $\mathcal{S}$ (and vice versa).

\subsection{Measurement probability (Proof of lemma~\ref{lemma:reverse})}

In the system-reference entanglement distillation algorithm, the probability of measuring $m$ and $\overline{m}$ is given by 
\begin{align}
 \figbox{1.5}{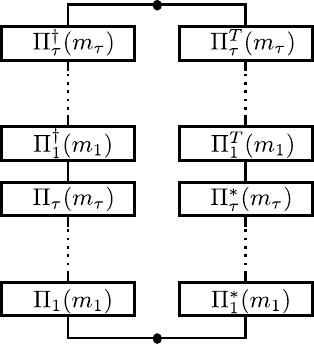}. 
\end{align}
Observe that this quantity is identical to $\Prob(m,\overline{m};I_{A})$ in Eq.~\eqref{eq:Prob(m,m,PA)} with $P_{\tau},\cdots, P_{1}$ arranged in the reverse chronological order. Then, from lemma~\ref{lemma:probability}, we can show that one may measure $s=m\cdot \overline{m}$ if and only if
\begin{align}
s \in \E_{\rev}.
\end{align}
This proves lemma~\ref{lemma:reverse}. 

\subsection{Entanglement distillation from reference (Proof of lemma~\ref{lemma:output-sys-ref-distillation})}

By treating the reference $R$ as a part of the system, one can utilize the algorithm from section~\ref{sec:distillation} to distill entanglement between $AB$ and $R$. The whole procedure is graphically summarized as follows:
\begin{align}
 \figbox{1.5}{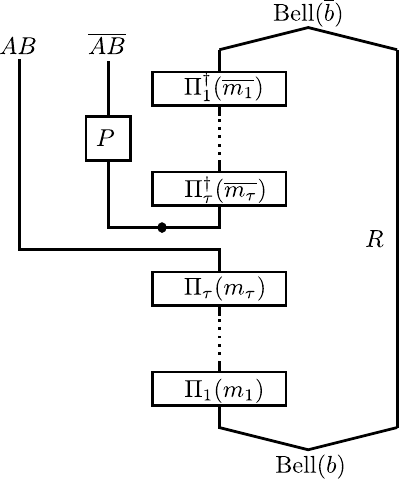} \  \label{eq:fig-AR-distillation}
\end{align}
where $\Bell^{\dagger}(\overline{b})$ represent the reverse Bell measurements with outcomes $\overline{b}$. Note that we can set $b = (1,\cdots, 1)$ by preparing EPR pairs on $ABR$ at the beginning, instead of performing Bell measurements. 

The distillation algorithm proceeds by finding an appropriate feedback Pauli operator. Let us introduce the following extended vectors:
\begin{align}
\overline{m}_{\ext} \equiv (\overline{m}, \overline{b}) \qquad s_{\ext} \equiv (s, b\cdot \overline{b}).
\end{align}
Then, after performing the aforementioned protocol, we compute $s_{\ext}$ and solve for $P$ satisfying
\begin{align}
s_{\ext} \in \E_{\ext}^{(P)}.
\end{align} 
By using lemma~\ref{lemma:output}, we notice that the output of the distillation algorithm, averaged over the measurement outcomes, is given by
\begin{align}
\mathbb{E}(\sigma_{AB\overline{AB}}) = \frac{1}{{N_{I}}_{\ext}} \sum_{ P \in \mathcal{S}} |P\rangle \langle P | \label{eq:SR-distilled}
\end{align}
where ${N_{I}}_{\ext}$ is the number of Pauli operators which satisfy $\C_{\ext}(P)\in \E_{\ext}$ (\emph{i.e.} the number of elements in $\mathcal{S}$). From this expression, one may see that $\mathcal{S}$ indeed plays the role of the stabilizer group. 

Finally, let us prove lemma~\ref{lemma:output-sys-ref-distillation}. Recall that the feedback operator in the aforementioned distillation algorithm reduces the system to the situations with $m=\overline{m}$ and $b=\overline{b}$. Also observe that this entanglement distillation in Eq.~\eqref{eq:fig-AR-distillation} is identical to the one from the main part of the paper in Eq.~\eqref{eq:MQC-summary-Ref} when $b=\overline{b}$. Hence, it suffices to prove that the feedback operation for the algorithm in Eq.~\eqref{eq:MQC-summary-Ref} reduces the system to the situations with $m=\overline{m}$.

This can be proven from the following observation:
\begin{align}
P_{j} \Pi(m) =P_{j} \frac{I + m_{\tau} P_{\tau}}{2} \cdots \frac{I + m_{j} P_{j}}{2} \cdots \frac{I + m_{1} P_{1}}{2} = m_{j}\Pi(m\cdot \E_{\rev}(P_{j})).
\end{align}
So, we have 
\begin{align}
P_{\Lambda} \Pi(m) \propto \Pi(m\cdot \E_{\rev}(P_{\Lambda})).
\end{align}
Hence, applying $P_{\Lambda}$ reduces the system to the situation with $m=\overline{m}$. Thus, the output of the distillation algorithm from section~\ref{sec:R-distillation} outputs the quantum state in Eq.~\eqref{eq:SR-distilled}. This proves lemma~\ref{lemma:output-sys-ref-distillation}.

\providecommand{\href}[2]{#2}\begingroup\raggedright\endgroup


\end{document}